# IT$_S$, A Structure Sensitive Information Theory

Samuel (Shmuel) Sattath
samuel@sattath.com

Hebrew University Jerusalem

3 July 2006

**Abstract**

Broadly speaking Information theory (IT) assumes no structure of the underlying states. But what about contexts where states **do** have a clear structure – how should IT cope with such situations? And if such coping is at all possible then – how should structure be expressed so that it can be coped with? A possible answer to these questions is presented here. Noting that IT can cope well with a structure expressed as an accurate clustering (by shifting to the implied reduced alphabet), a generalization is suggested in which structure is expressed as a measure on reduced alphabets. Given such structure an extension of IT is presented where the reduced alphabets are treated simultaneously. This structure-sensitive IT, called $IT_\mathbf{s}$, extends traditional IT in the sense that: a) there are structure-sensitive analogs to the notions of traditional IT and b) translating a theorem in IT by replacing its notions with their structure-sensitive counterparts, yields a (provable) theorem of $IT_\mathbf{s}$. Seemingly paradoxically, $IT_\mathbf{s}$ extends IT but it's completely within the framework of IT. The richness of the suggested structures is demonstrated by two disparate families studied in more detail: the family of hierarchical structures and the family of linear structures. The formal findings extend the scope of cases to which a rigorous application of IT can be applied (with implications on quantization, for example). The implications on the foundations of IT are that the assumption regarding no underlying structure of states is not mandatory and that there is a framework for expressing such underlying structure.

# Introduction

Information theory (IT) applies to states[1] on which a probability is defined. Nothing more is required by the theory and, particularly, an underlying structure of the states is not a premise.[2] However, in some contexts states do have an underlying structure and it is easy to show that the structure must affect IT (see section 1.1). This raises a question regarding the way IT can cope with a structure of states . That is the subject of this manuscript.

Since structure of states has no direct role in IT, structure must be defined before being brought to the framework of IT. This work presents a formalization of the concept of structure so that it can be directly accommodated by an extension of IT. The extended IT is presented in two stages (see Figure 1.0.1). Part 1 lays the ground. It focuses on hierarchical structures and formulates a structure sensitive expression for two key notions: entropy and code length. These formulations are utilized in Part 2 to suggest a broad class of structures to which significant parts of IT can be extended to form a structure sensitive IT.

Part 1 starts with a familiar case where IT **can** cope with structure of states: the case where states are accurately portrayed by clustering. In such case IT has a recipe for coping with the structure by shifting to a reduced alphabet. This recipe is extended to address the case where the structure is hierarchical (i.e. equivalent to an ultrametric distance), leading to an axiomatization of distance-sensitive entropy for ultrametric distances ($H_U$). $H_U$ is shown to have a formulation where the structure is expressed as a measure on a set of partitions. It is also shown that in an ultrametric context $H_U$ is the bound on code length, when code length is redefined to take into account distance between letters.

Part 2 is the heart of this work. It unveils a general class of structures, for which a structure-sensitive extension of IT ($IT_S$) is possible. The formalization of structure is inspired by the previously mentioned formulation of $H_U$ where structure is expressed as a measure on partitions (or, equivalently, reduced alphabets). For such structures there is a translation procedure that converts the most basic notions in IT to their structure sensitive counterparts so that elementary theorems in IT are retained in $IT_S$. Following that, other aspects of IT (typical sequences and coding) are studied: again, it is shown that the structure-sensitive analogs of theorems in IT are valid in $IT_S$. The last section of Part 2 is a discussion of the findings that, broadly speaking, $IT_S$ extends and generalizes IT.

Part 3 applies $IT_S$ to the case where states are embedded on the real line ($IT_R$). This is an interesting structure since numerous applications have states in this form. Some of the unique features of $IT_R$ are presented, and the implication on quantization is mentioned.

The following table presents the scope of each part:

| Theory | traditional IT | $IT_U$ | $IT_S$ | $IT_R$ |
|---|---|---|---|---|
| structure | uniform | ultramertric | partition | Linear |
| Partition structure | singletons | hierarchical | any | real numbers |
| Part | | 1 | 2 | 3 |

---

[1] The terms 'state' and 'letter' are used interchangeably.
[2] Formally, this is tantamount to assuming a uniform structure on states, like a Hamming distance.

# Part 1 – Extending IT to accommodate a hierarchical structure

Part 1 starts with demonstrating a problem with the notion of entropy: its deficiency in accommodating a case where the states under discussion have a clustering structure. Then a remedy for this specific case is shown: shifting to a reduced alphabet. Following that, a broader class of structures is presented – the class of hierarchical structures which is formalized as the class of ultrametric trees. It is shown that the remedy of shifting to reduced alphabets extends to hierarchical structures. To do that an axiomatic formulation of ultrametric entropy ($H_U$) is shown and studied. The last sections of Part 1 discuss coding in an ultrametric context, and present applications of $H_U$.

Part 1 prepares the ground to Part 2. Part 2 generalizes and expands Part 1 and implies that an ultrametric extension of IT ($IT_U$) is possible. The relation between the two parts is presented in Figure 1.0.1.

| Part | 1 | 2 |
|---|---|---|
| Structure | Hierarchical | Partition Structures |
| Theory | $IT_U$ | $IT_S$ |
| Parts of information theory that are handled | Entropy | Basic Notions [3] |
| | | Typical Sequences |
| | Coding | Grouping |
| | | Coding |

Figure 1.0.1 presents the relation between Part 1 and Part 2. Part 1 covers hierarchical structures – just one family in the class of structures studied in Part 2. The reason to bother specifically with hierarchical structures in Part 1 is that two findings in Part 1 guide us in Part 2 (see **brown arrows**):
1. The structure of weights on partitions appears in 1.3.7 and 1.3.9. This is taken as the definition of structure in Part 2. It also transfers directly to Part 2 and underlies the translation procedure of section 2.2 (see 2.2.1.), which establishes in $IT_S$ the notion of entropy and all other basic notions.
2. The definition of code length in the ultrametric context (see 1.4.2) introduces distance into code length. This guides us in Part 2 in handling the grouping property and coding.
One could have shown that other aspects of IT can cope with hierarchical structures, thus establishing an ultrametric extension of IT, namely $IT_U$. This would be futile since it follows directly from Part 2 where a stronger result is presented – that $IT_S$ extends traditional information theory, and since $IT_U$ is a special case of $IT_S$ (as indicated by the **green arrow**).

## Section 1.1 A deficiency with the notion of entropy and a partial solution

Entropy is widely used as a measure of 'randomness' [4]. As such there are some intuitive constraints it must abide to. The synthetic example presented in Figure 1.1.1.a demonstrates a deficiency with entropy as a notion of randomness when states have a hierarchical structure. Four letters, $\alpha$-$\delta$, are portrayed as the leaves of a tree so that the distance between any two letters is the length of the path connecting them. The letters fall into two clusters: {αβ} and {γδ}. The inter cluster distances are, say, 1 and the intra cluster distances in Figure 1.1.1.a are negligible, say 0. Compare the case where the probability of states is P: P(α)=P(β)=0.5 (and other letters have probability 0) with the case where the probability is Q: Q(α)=Q(γ)=0.5. For the distances portrayed in Figure 1.1.1.a, P is clearly less random than Q since the letters α and β are almost indistinguishable as far as the distance is concerned whereas α and γ are not. Thus the two probabilities P and Q, which have the same value of randomness when distances are as depicted in Figure 1.1.1.c, have different values of randomness when distances are as in Figure 1.1.1.a. The conclusion is that in this context there is a deficiency with entropy as a measure of randomness, since it ignores the distances between states.

---

[3] The basic notions are : entropy, entropy of joint probability, conditional entropy, mutual information and relative information
[4] This informal use of 'randomness' is intentionally vague.

However, there is a way around this deficiency in the case of the 4-letter alphabet of Figure 1.1.1.a: this alphabet is reducible to a 2-letter alphabet in which each cluster is one letter. When shifting to the reduced alphabet (see section 2.1 for a formal treatment), the randomness in Figure 1.1.1.a can be measured by the entropy of the reduced alphabet.

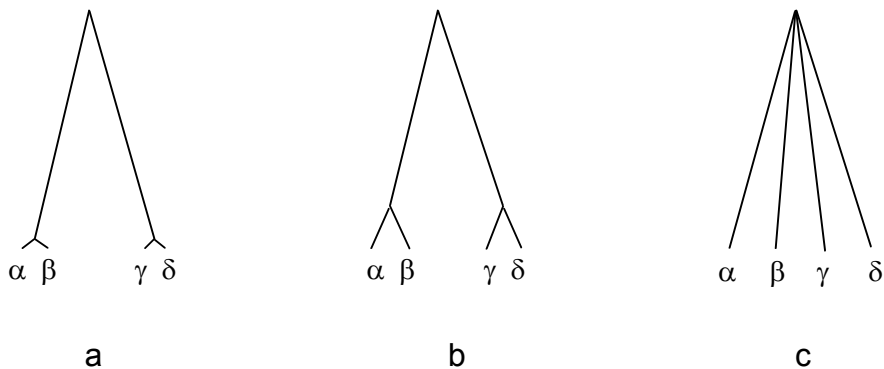

Figure 1.1.1. Trees portraying distances between four letters: the distance between a pair of letters (leaves) in a tree is the length of the shortest path connecting them. See text for details.

The use of reduced alphabets is a known practice in applying IT to cases where states have a structure that is well portrayed by clustering. A non-synthetic example of applying reduced alphabets is found in biology, see 1.5.1.

A clustering of states is an underlying structure that IT **can** cope with: all that is required in order to apply IT is a transition from the original alphabet to the reduced alphabet. In the rest of Part 1 the logic of coping with reduced alphabets will be extended to coping with hierarchical structures. In Part 2, a further generalization of reduced alphabets will be presented and shown to enable extension of IT to a wide class of structures.

### Section 1.2 Basic definitions and notation

This section presents some necessary definitions and notations.

**Definitions 1.2.1.** (probabilities and distances): Let A denote a finite set of states also referred to as letters and, collectively, as alphabet. Let P denote a probability on A where P(a) is the probability of a ε A and P(B) is the probability of B ⊆ A. Let D denote a distance between states.

1. For any two disjoint non-empty sets of states $B,C \subseteq A$, $D(B,C)$ is defined to be the expected distance between states in B and states in C, i.e.

$$D(B,C) = \sum_{b \, \varepsilon \, B} \sum_{c \, \varepsilon \, C} P(b|B)*P(c|C)*D(b,c)$$

2. For any set of states B, D|B denotes the distance D restricted to states in B.
3. Let $B \subseteq A$. $Y=\{B_1 \ldots B_k\}$ is a partition of B if all $B_i$ are non-empty, disjoint, proper subsets of B and $UB_i=B$. The $B_i$'s are called components of Y.
4. Let $B \subseteq A$ and let $Y=\{B_1 \ldots B_k\}$ be a partition of B.
   a. P induces a probability on Y, denoted by $P^Y$ and defined by $P^Y(B_i)=P(B_i|B)$.
   b. Similarly, D induces a distance on Y denoted $D^Y$ and defined by $D^Y(B_i,B_j)=D(B_i,B_j)$ (the expression on the right is the distance defined in 1. above).

**Definition 1.2.2.** (trees and their composition):
1. Trees are finite, acyclic, undirected, non-negatively weighted, connected graphs with a designated node of degree > 1 called root.
2. The leaves of a tree are the nodes with degree 1.
3. The tree-distance between two nodes is the length of the path connecting them.
4. Given a tree T, for any non-root node $i$ in T there is a unique node adjacent to $i$ in the path connecting $i$ to the root. It is called the parent of $i$ and denoted Parent($i$).
5. Given a tree T, a node $i$ in T is defined as a descendant of a node $j$ if $j$ resides on the shortest path connecting $i$ to the root. Thus a node is a descendant of itself. We say $i$ is an immediate descendant of $j$ if Parent($i$)=$j$.
6. Any node on the path from a node $i$ to the root, excluding $i$, is an ancestor of $i$.
7. Any two nodes have a common ancestor (e.g. the root). There is only one node that is their least common ancestor – it is defined as the node that has no descendant that is a common ancestor.
8. For any node $i$ in T, $T_i$ will denote the subtree in which the nodes are the descendants of $i$, $A_i$ will denote the set of leaves of $T_i$.
9. For any node $i$ in T, $\{A_j: j$ is a direct descendant of $i\}$ is a partition of $A_i$ called the natural partition of $A_i$ and denoted by $Y_i$. $Y_{root}$ is called the natural partition of T.

**Definition 1.2.3.** Let T be a tree with a set of leaves A on which there is a probability P. For any node $i$ in T, define $P_i=P(A_i)$, i.e. the probability of the leaves that are descendants of $i$.

Informally, a hierarchical structure is the result of taking a set of states, partitioning it into clusters, then further partitioning each cluster and continuing recursively until all clusters are singletons. A good way to formalize that is to use ultrametric distances:

**Definition 1.2.4.** (ultrametric distance, see (6)):
1. An ultrametric distance D on a set A is a distance satisfying the following ultrametric-inequality:  for any $a, b, c \, \varepsilon \, A$: $D(a, b) \leq \max(D(a,c),D(b,c))$.
2. An (ultrametric) distance where D is constant is said to be uniform.
3. An ultrametric distance with maximal distance of 1 is called normalized.

The interesting property of ultrametric distances is their well known equivalence with the class of trees in which all leaves have the same distance from the root. Take any tree with leaves that are equidistant from the root, as depicted in Figures 1.1.1 and 1.4.1. It is easy to see that in such trees the tree-distance between pairs of leaves is ultrametric. The reverse is also true and is presented without proof:

**Theorem 1.2.5.**: Given an ultrametric distance D on a set A, there exists a unique tree, denoted $T_D$, with leaves A so that the tree-distance between any two leaves is D. In $T_D$ the tree-distance between the root and the leaves is constant. Moreover

1. It is possible to assign to every node $i$ in $T_D$ a non-negative real number height($i$) so that the distance between any two leaves is the height of their least common ancestor.
2. Assignment of height to non-leaves is unique. It is natural to assign leaves height=0.
3. If the ultrametric tree is normalized then the height of the root is 1.
4. Let $i$ be an internal node in $T_D$ and let $Y_i$ be the natural partition of $A_i$ (see 1.2.2.9). Let $B_1$ and $B_2$ be two different components of $Y_i$ (see 1.2.1.3). Then the following holds: $\forall b_1 \varepsilon B_1 \forall b_2 \varepsilon B_2 \ D(b_1,b_2)$ = height($i$).

**Definition 1.2.6.** (ultrametric trees): Given an ultrametric distance D,

1. $T_D$ in 1.2.5 is defined as the (ultrametric) tree induced by D.
2. An ultrametric tree is a tree induced by an ultrametric distance. The terms ultrametric distance and ultrametric tree will be used interchangeably.
3. For any node $i$ in an ultrametric tree T, define $L_i$ = height(parent($i$)) – height($i$), where height($j$) is as defined in 1.2.5.1. (it is easy to show that $L_i$ is twice the length of the arc that connects $i$ to parent($i$) in T).

## Section 1.3 Entropy on a hierarchical structure

**Definition 1.3.1.** (notation for entropy): Entropy is denoted as $H(P)$, indicating its dependence solely on probability. It is shortened to H when context permits. Its distance-sensitive generalization to ultrametric distances, to be introduced shortly, will be denoted by $H_U(P, D)$ or, context permitting, $H_U$.

Entropy is well known to have an axiomatic formulation where the main axiom is grouping (see (1), p. 43). The ultrametric entropy $H_U$ can also be formalized axiomatically. In case that the distance does not matter we would want $H_U(P, D)$ to reduce to the traditional entropy. This means that if D is uniformly 1 then $H_U$=H. When D is uniform but different than 1, the distance D will be considered as a scaling constant. What remains is to define $H_U$ when D is not constant. Consider the distances depicted in Figure 1.1.1.b: there are two clusters of letters ($\alpha\beta$ and $\gamma\delta$) with the inter-cluster distances considerably bigger than the non-zero intra-cluster distances. Within each cluster the distances are uniform so we already know how to compute the intra-cluster entropy. However, if the intra cluster entropy is known it can be removed, and that yields a reduced alphabet in which the distance is also uniform. So on the reduced alphabet we also know how to calculate $H_U$. This resolves the question of calculating $H_U$ in the case of Figure 1.1.1.b. Using recursion, the same principle can be applied to any ultrametric tree. This leads to the following recursive formulation of $H_U$:

**Axiomatization 1.3.2.**: Given an ultrametric distance D and a probability P on a finite set A, let T be the tree induced by D. $H_U(P,D)$ is defined recursively on T as follows:

a. Homogeneous extension.

If the distance D is constantly d (i.e. $\forall a,b: D(a,b) = d$) then $H_U$ is homogeneous with H:
$$H_U(P, D) \equiv d * H(P)$$

b. Grouping axiom for the natural partition

Let $Y = \{A_1 \ldots A_k\}$ be the natural partition of T (see 1.2.2.9). Then
$$H_U(P, D) = H_U(P^Y, D^Y) + \sum_{i \text{ is a direct descendant of the root}} P(A_i) * H_U(P|A_i, D|A_i)$$

Where $P^Y$ and $D^Y$ are as defined in 1.2.1.4 and $A_i$ is as defined in 1.2.2.8

It is easily confirmed that the distances $D^Y$ and $D|A_i$ above are ultrametric, thus validating the appropriateness of the recursion in 1.3.2.b.

When comparing the axiomatic formulation of $H_U$ to the traditional formulation of H one sees that the technical axioms (normalization and continuity, see (1), p.43) are not necessary since they are covered by the extension axiom 1.3.2.a. The traditional grouping rule applies but is restricted only to the natural partition [5]. The full grouping property must be given up since if it were to apply to all partitions then $H_U$ would be H.

Like in traditional IT, the recursive formulation of entropy is not very useful. In attempting to find a non recursive formulation for $H_U$ it is clear that the key is the term $H_U(P^Y, D^Y)$ in 1.3.2.b . We will now see how this term is calculated.

**Lemma 1.3.3**: Given an ultrametric distance D on a finite set A and a probability P on A, let T be the tree induced by D and let Y be the natural partition of T (see 1.2.2.9) then
$$H_U(P^Y, D^Y) = H(P^Y) * \text{height(root)}.$$

Where $P^Y$ and $D^Y$ are as defined in 1.2.1.4 and height is as defined in 1.2.5.1

**Proof**: The distance $D^Y$ is constant – its value is height(root) (see 1.2.5.4), thus lemma 1.3.3 results from the homogeneity axiom 1.3.2.a.

The recursive formulation of $H_U$ (1.3.2) has an equivalent non-recursive formulation:

**Theorem 1.3.4**: Given an ultrametric distance D on a finite set A and a probability P on A, let $T_D$ be the tree induced by D. Then
$$H_U(P, D) = \sum_{i \text{ is a non-leaf node in T}} P_i * \text{height}(i) * H(P^{Y_i})$$

**Proof**: The proof uses Lemma 1.3.3. and it is by recursion on the ultrametric tree. It appears in the Appendix.

---

[5] A revised grouping property that applies to all partitions will be introduced in section 2.5

**Theorem 1.3.5**: Given an ultrametric distance D on a finite set A and a probability P on A, let $T_D$ be the tree induced by D (see 1.2.5 and 1.2.6).

Then [6]        $H_U(P,D) = -\sum\limits_{\substack{i \text{ is a non-root node in } T_D}} L_i * P_i * log(P_i)$

**Proof**: The proof appears in the Appendix.

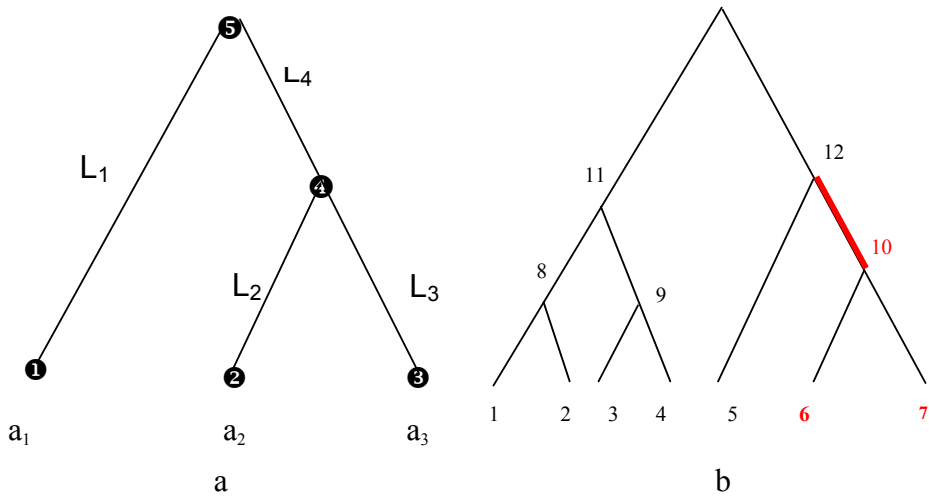

Figure 1.3.1. Examples of computing $H_U$ according to formulation 1.3.5

**a**.  For the 3-leaf ultrametric tree on the left,
$-H_U(P, D) = L_1 * P(a_1) * log(P(a_1)) + L_2 * P(a_2) * log(P(a_2)) + L_3 * P(a_3) * log(P(a_3))$
$+ L_4 * (P(a_2) + P(a_3)) * log(P(a_2) + P(a_3))$

**b**.  To calculate $H_U$ for the 7-leaf ultrametric tree on the right, sum $L * P * log(P)$ for 12 nodes. For example, node 10 with length L drawn in red contributes the expression $L * P * log(P)$ where P is the probability of its two underlying (red) leaves.

Theorem 1.3.5 highlights a noteworthy similarity of $H_U$ to H:

$$H_U(P,D) = -\sum\limits_{\substack{i \text{ is a non-root node in } T_D}} L_i * P_i * log(P_i) \qquad\qquad H(P) = -\sum\limits_{\substack{i \text{ is a leaf in } T_D}} P_i * log(P_i)$$

both are calculated as a sum of expressions of the form $L * P * log(P)$. In the traditional case this is computed only for leaves and $L \equiv 1$. For $H_U$ the computation ranges over all nodes of an ultrametric tree, (leaves as well as internal nodes) see Figure 1.3.1. Moreover, on a normalized trivial tree (e.g. Figure 1.1.1.c) the first expression reduces to the second.

---

[6]  The range of the index in this equation can be expanded to include the root by defining $L_{root}$ to have any value.

There is yet another formulation of $H_U$ that will prove valuable further down the line (see Figure 1.3.2). Imagine a continuous process that starts, at height 1 (at the root of the tree) and traverses from the root towards the leaves in a constant velocity ending at a leaf (at height 0). When this process encounters an internal node it chooses one of the branches according to the (conditional) probability of the leaves at the bottom of that branch. The probability that this process terminates at a given leaf is exactly the probability of the leaf and the probability that it passes through any arc is the probability of the leaves falling under the arc. The probability of the process being exactly on a node is zero so one can assume, for convenience, that the process is always on one of the arcs. At any given point in time, t, the process is at a fixed height and it can be on several arcs. Each of these arcs has some of the leaves under it. The sets of leaves under these possible arcs form a partition of leaves at time t, call it $s_t$. The partitions $s_t$ become finer as the process approaches the leaves. Each of the components of $s_t$ has a probability (the sum of the probabilities of the leaves). Using that probability, it is possible to compute the traditional entropy of $s_t$, denoted by $H(s_t)$. The integral of $H(s_t)$ over time is $H_U$. A formal statement of this fact follows.

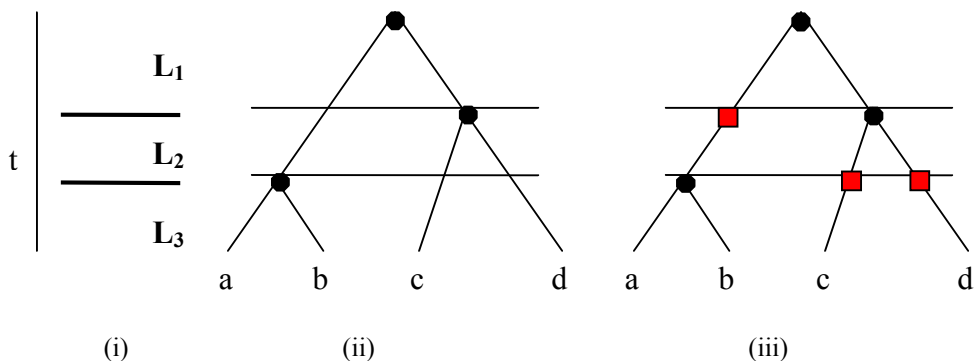

Figure 1.3.2. An explanation of the formulation of $H_U$ presented in 1.3.7-1.3.9. (ii) Portrays a normalized ultrametric tree. (i) Presents the vertical scale of height, t, which goes from zero at bottom to 1 at the top. The horizontal lines drawn at the height of each of the internal nodes partition the vertical scale t into three bands. (iii) Presents the banding of the tree in (ii), i.e. the addition of internal nodes (red squares) where band boundaries intersect the tree. Each band partitions the leaves of the tree and the lower the partition the finer the partition. $H_U$ is calculated by computing the traditional entropy for each band and adding these entropies weighted by the height of the band.

**Definition 1.3.7.** (bands and their measure in an ultrametric tree): Let T be a normalized ultrametric tree with leaves A and probability P on the leaves. (see Figure 1.3.2).
1. The tree T is **banded** if for every node x and leaf a, on the path from a to the root there is a node y with height(y)=height(x). In Figure 1.3.2 (iii) is banded and (ii) is not.
2. For any $0 \leq t \leq 1$, define  $r_t = \{i : i \text{ is a node in T and height}(i)=t\}$.
   Clearly $r_t \neq \phi$ iff there is a node at height t. If $r_t \neq \phi$, t will be called realizable.
3. For a banded tree and a realizable t let $s_t$ be the following partition of leaves
   $$s_t = \{A_i : i \; \varepsilon \; r_t\} = \{A_i : i \text{ is a node in T and height}(i)=t\}.$$
   and let the entropy of the partition be defined as

$$H(s_t) = - \sum_{i \, \varepsilon \, r_t} P_i * \log(P_i)$$

(A$_i$ was defined in 1.2.2.8 and P$_i$ in 1.2.3.)

4. For any $i$ and $j$ in $r_t$, height($i$)= height($j$). Therefore, for a banded tree and a realizable $t < 1$ the height of the band at $t$ can be defined as:

$\hat{S}(t) = $ height($i$) − height(parent($i$))    where $i$ is in $t$.

(i.e. the value of $\hat{S}(t)$ does not depend on the choice of $i$ as long as it is in $t$)

5. For a banded tree: S={$s_t$: $t$ is realizable} is a set of partitions with a measure $\hat{S}(t)$.

**Lemma 1.3.8.** (indifference to banding):

1. Given an ultrametric tree T with leaves A apply the following procedure: for any internal node $x$ and any leaf $a$, add a node $y$ on the path from the root to $a$ so that height($y$)=height($x$) (if such a node does not already exist). This procedure yields a banded tree, referred to as the 'banding' of T and denoted T$_B$.

2. Given an ultrametric tree T with leaves A and probability P on leaves,

$$H_U(P,D) = - \sum_{i \textbf{ is a non-root node in T}} L_i * P_i * \log(P_i)$$

is equivalent to

$$H_U(P,D) = - \sum_{i \textbf{ is a non-root node in T}_B} L_i * P_i * \log(P_i)$$

Proof: Immediate.

The following theorem expresses H$_U$ as a weighted average of traditional entropies calculated on partitions induced by the tree. This formulation establishes a relation between a structure sensitive notion (H$_U$), its traditional counterpart (H) and weights on partitions ($\hat{S}$). Such relationship will be key in Part 2.

**Theorem 1.3.9.** (formulating H$_U$ as a weighted sum of traditional entropies of partitions):

$$H_U = \sum_{t \textbf{ is realizable}} \hat{S}(t) * H(s_t)$$

**Proof**: The proof appears in the Appendix.

Up till now we discussed in detail ultrametric entropy. The broader claim that IT can be extended to accommodate ultrametric distances is deferred to Part 2 as explained in Figure 1.0.1. Two results of Part 1 are required for Part 2: one of them is 1.3.9 and the other is a formalization of distance-sensitive coding, to which we turn now.

**Section 1.4 Ultrametric coding**

This section shows that the appealing property of entropy being a lower bound of code length extends to ultrametric contexts. $\mu^c$ will denote the traditional average-code-length of a code tree C and $\mu^c_U$ will denote its distance sensitive analogue for ultrametric trees. We start with the motivation leading to the definition of $\mu^c_U$.

Consider an alphabet with distances as portrayed in Figure 1.4.1.1 and with a uniform probability. The uniform probability implies that all binary balanced code trees (Figures 1.4.1.2, 1.4.1.3 and 1.4.1.4) are equally effective in coding. However, when distances are considered, this symmetry of code trees breaks **during** the process of decoding, in the following sense. A decoding process can be viewed as a gradual revelation of a transmitted letter, each node in the code tree representing a range of possible values that are still open for the transmitted letter to have. An internal node of a binary code-tree partitions the possible letters according to the next bit being 0 or 1. Call these sets of letters $A_0$ and $A_1$ respectively. We assign to each internal node the value $D(A_0,A_1)$ (see 1.2.1.1) and imply that the bigger this distance the bigger the resolution power of the node. From the three possible code trees in Figure 1.4.1, the one that has the same topology as the ultrametric tree (namely Figure 1.4.1.2) has the biggest resolution power in its root: once the first bit was received the transmitted letter is known up to a small distance. In analogy to $\mu^c$ being the expected number of nodes traversed in C, we will define $\mu^c_u$ to be the expected resolution power. When distances are constant the resolution power is also constant and $\mu^c_u$ reduces to $\mu^c$. We will now formally define $\mu^c_u$, together with a related value, $\lambda^c_u$, needed in its investigation.

**Definition 1.4.1.**: Let h(p) be the function defined for any $0 \leq p \leq 1$ as
$$h(p) = - (p * \log (p) + (1-p) * \log (1-p)).$$
Given two disjoint sets of letters $A_0$, $A_1$ in an alphabet with probability P we also define
$$h (A_0,A_1) = h (P(A_0|A_0 \cup A_1))$$

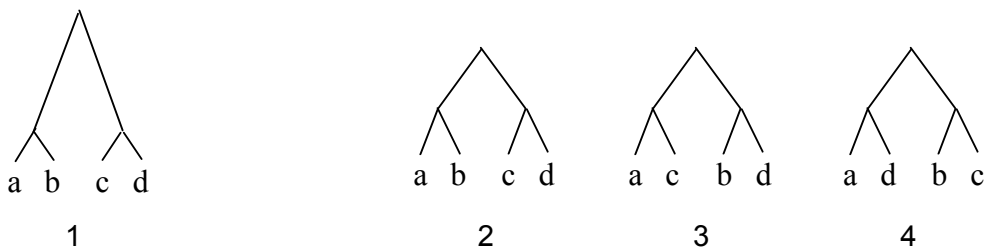

a b c d        a b c d    a c b d    a d b c

   1             2         3         4

Figure 1.4.1. An alphabet of four letters has distances portrayed in **1** above. Each letter has a probability of 0.25. The three possible binary balanced code trees are presented in **2**, **3** and **4**. Only code tree **2** has the same topology as **1**.

**Definition 1.4.2.**: Let C be a binary code tree on an alphabet A. We use recursion on the code tree C, to define $\mu^c_u$ and $\lambda^c_u$.

a. If A has more than one letter, let $C_0$, $C_1$ be the two sub-trees stemming from the root and let $A_0$, $A_1$ be their respective leaves.

$$\mu^c_u (A, P, D) = D (A_0, A_1) + \sum_{i=1,2} P (A_i) * \mu^{ci}_u(A_i, P|A_i, D|A_i)$$

$$\lambda c_u(A, P, D) = h(A_0, A_1) * D(A_0, A_1) + \sum_{i=1,2} P(A_i) * \lambda c i_u(A_i, P|A_i, D|A_i)$$

b.  If the alphabet A has only one letter we set:  $\mu c_u(A, P, D) = \lambda c_u(A, P, D) = 0$.

Note the analogy with the traditional definition: $\mu c(A, P) = 1 + \sum_{i=1,2} P(A_i) * \mu c i(A_i, P|A_i)$.

The cumbersome notation $\mu c_u(A, P, D)$ is often simplified to $\mu c_u$ when the alphabet, probabilities and distance are obvious from the context and to $\mu_u$ when the code tree is also obvious.

**Theorem 1.4.3.**: For any ultrametric D with leaves A, a probability P on A and a binary code tree C on the alphabet A: $H_u \le \mu c_u$.

Theorem 1.4.3, proven in the appendix, is the analogue of the traditional statement that average code length cannot fall from the entropy.

The analogue to the traditional reverse of theorem 1.4.3 is that for any normalized ultrametric distance D there exists a binary code C so that that $\mu c_u(A,P,D) \le H_u(P,D) + 1$.

**Algorithm 1.4.4 Compressed coding algorithm for $H_u$**. There is an algorithm that given an ultrametric distance with leaves A and a probability P on A, produces a binary code tree with $\mu c_u \le H_u + 1$.

The details of the algorithm are presented in the Appendix. No analytic proof of its validity is presented, but extensive testing of the algorithm failed to produce any counter-example.

To summarize, we have seen that the ultrametric analogue to traditional code length is a notion of distance-sensitive code length. More on that in Part 2.

## Section 1.5 Applications of ultrametric IT

We have seen only two distance-sensitive notions (entropy and coding) for ultrametric distances. However, as explained in Figure 1.0.1., traditional IT can be extended to a theory, $IT_u$, that is sensitive to ultrametric distances and encompasses many notions and theorems of IT.  One way to assess whether such extendibility of IT is more than an anecdotal formal curiosity is to examine its applications. One could argue that the applicability of $IT_u$ is restricted by the limited scope of ultrametric distances. To demonstrate that this is not too severe to bar practical applications, two applications of $H_u$ will now be presented.

### 1.5.1 Measuring positional protein conservation

Measuring positional conservation of amino acids in protein sequences is an important procedure in the study of proteins. The reason is that high conservation in a position (column) generally implies that the amino acid (AA) in that position is essential for the protein's structure and/or function (9-14). In order to utilize this empirical phenomenon and identify conserved positions, a score of positional conservation is required. The calculation of conservation score starts with portraying proteins, which are polypeptide chains, as a sequence of the amino acids (AAs) that compose them. This yields a sequence in the 20 letter alphabet of AAs. Proteins that belong to the same family are aligned so that they form a multiple-sequence-alignment (MSA) presented as a matrix where each row is a protein and each column is a specific position (see Figure 1.5.1.). To assess the level of conservation in a given position one can calculate the frequency of each letter in that position and compute the entropy of the resulting probability. Since AAs are well known to have different biochemical and physical properties, using entropy as a measure of conservation is deficient in that it does not take the similarity of AAs into account. Among the many methods that were employed to calculate conservation while accounting for this inherent similarity, one approach is to group the AAs according to their biochemical properties into a small number of clusters and calculate the entropy of the resulting reduced alphabet.

Given a MSA, numerous conservation scores have been suggested (for a comprehensive review see (15)). Such measures are typically not based on a formally grounded approach for combining the frequency/probability of AAs with their similarity into one measure of conservation. To demonstrate the relevance of our formal findings we turn now back to the question – how should such positional conservation be measured from a MSA?

```
Q59265/26-255       IFQVATIGSRAR.CLRWRFYYGKLQKKGNFGLGTFLDLNGEMVAVDGHYYEIEAN
Q9Z632/3-220        LFQHSTMAALVGGLFSGTTSFKELLQHGDLGITLDQFDGELIILDGEAYKVKPE
ALDC_BACSU/12-242   IYQVSTMTSLLDGVYDGDFELSEIPKYGDFGITFNKLDGELIGFDGEFYRLRSD
Q53405/67-293       LYQTSTMAALLDAVYDGETTLDELLMHGNFGLGTFNGLDGEMIVNDSVIHQFRAD
ALDC_KLETE/19-249   IYQTSLMSALLSGVYEGSTTIADLLTHGDFGLGTFNELDGELIAFSSEVYQLRAD
ALDC_BACBR/46-272   LFQYSTINALMLGQFEGDLTLKDLKLRGDMGLGTINDLDGEMIQMGTKFYQIDST
ALDC_LACLC/1-227    LFQYNTLGALMAGLYEGTMTIGELLKHGDLGTLDSIDGELIVLDGKAYQAKGD
P94897/1-230        AYQHGTLAQIMDGQYDGTILLKDLLEHGDFGIGTTTGIGVELIVLDGVAYGIPSS
```

Figure 1.5.1. Multiple sequence alignment (MSA) of positions 11-65 for a subset of proteins from the AAL_decarboxy PFAM family (16). Amino acids are denoted by their one-letter code. The fully conserved positions are presented in bold. Assessing the level of conservation of the other positions requires a measure of conservation.

Clearly $H_U$ can serve as a measure of positional AA conservation **if** the distances between amino acids are ultrametric. A recent study (2) has shown that the distances implied by the widely used BLOSUM substitution scores can be well approximated by an ultrametric distance: BLOSUM scores can be approximated to an accuracy > 0.9 by an ultrametric structure (the accuracy of reconstruction being measured by Pearson correlation). Additional supporting evidence for the quality of the reconstruction is the stability of the structure across versions of BLOSUM and the fact that the structure reflects well known physicochemical properties.

## 1.5.2 Ultrametricity in the spin glass model

In physics, the spin glass model revealed an underlying ultrametricity that required a new formulation of entropy. The calculation of entropy performed in (17) was a special case of $H_U$ where the ultrametric tree is restricted to have two layers. The framework of $H_U$ enables to remove that constraint on the number of layers.

# Part 2 – structure dependent extension of IT

Part 2 builds on two findings of Part 1 as explained in Figure 1.0.1. Section 2.1 formalizes the concept of structure on states, which is used to develop a structure-sensitive IT (IT**s**). Sections 2.2-2.7 translate notions in IT to their structure-sensitive analogs and prove that the translations of theorems in IT are true (theorems in IT**s**). Finally, section 2.8 discusses the sense in which IT**s** extends IT. The formal arguments are rather straightforward though the notation is laborious.

### Section 2.1 A measure on partitions

The following definition formulates the concept of structure used in this study:

**Definition 2.1.1.** (structure): Given a set A,

1. A **partition-structure** on A is a pair (S, Ŝ) where S is a set of partitions of A, and Ŝ is a measure on S. Since Ŝ implies S, such Ŝ will also be referred to as a partition structure.
2. Ŝ is **normalized** if the measure sums to 1. In most cases, unless explicitly stated, Ŝ will not be assumed to be normalized.
3. A partition s **separates** states a and b if a and b are not in the same component of s. A set of partitions S is said to be **separating** if for any two states in A there is a partition in S that separates them. Ŝ is **separating** if for any two states in A there is a partition s in S that separates them and Ŝ(s)>0. Unless otherwise stated, it will be assumed that Ŝ is separating.[7]

The set A on which a partition-structure is defined will typically be a set of states on which a probability is defined. However, as far as the partition-structure itself is concerned, no such probability is involved. This separation of structure from probability is essential.

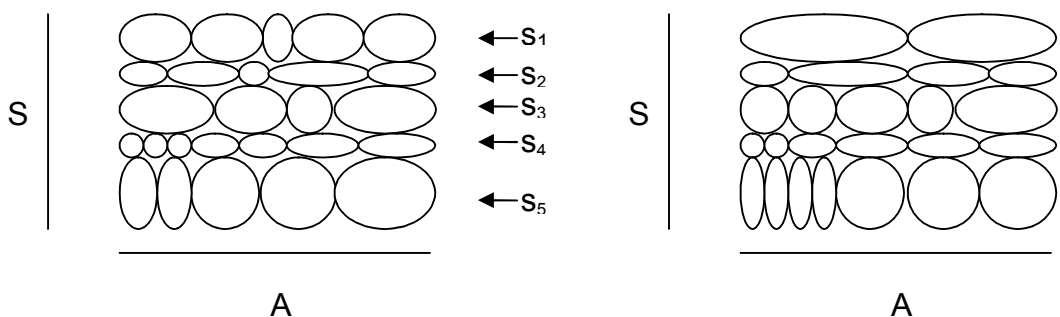

Figure 2.1.1. Two examples of S, each presenting 5 partitions of A, one partition in each row. Each ellipse is a subset of A. The width of an ellipse represents its probability P. The height of each row, s, represents Ŝ(s). Q, defined in 2.1.6, is the volume of each ellipse. The structure on the right is a hierarchical structure.

---

[7] There is no loss of generality in this assumption since two states that are not separated by any s in S are indistinguishable and can be united into one state.

**Examples 2.1.2** (examples of partition-structures):

1. Assume an ultrametric distance U between states A with maximal distance 1 (see Figure 2.1.2.). For any positive real number $r \leq 1$ and any $a \varepsilon A$ let $s_{a,r}$ be defined as $s_{a,r} = \{b: b \varepsilon A$ and $U(a,b) \leq r\}$. Let $s_r$ be the partition of A defined by $s_r = \{s_{a,r}: a \varepsilon A\}$[8]. $S = \{s_r: 0 < r \leq 1\}$ is a set of partitions. For any $0 < r \leq 1$ let $R_r = \{x: 0 < x \leq 1$ and $s_r = s_x\}$. Assigning to $s_r$ the measure $\max(R_r) - \min(R_r)$, yields a normalized measure $\hat{S}$ on S.[9]

2. Assume that the states A are points in the real interval [0,1]. Present A as $A = \{a_i\}_{i=1,...,n}$, where $a_i < a_j$, for $i < j$. Assume also that $a_1 = 0$ and $a_n = 1$ (see Figure 2.1.3.) For any $i < n$ let $s_i$ be $\{\{a: a \varepsilon A, a \leq a_i\}, \{a: a \varepsilon A, a > a_i\}\}$, a partition of A. S is defined as $\{s_i: i < n\}$ and the measure $\hat{S}$ of each $s_i$ is $a_{i+1} - a_i$.

3. The partition structure for traditional IT consists of a single partition, s, composed of singletons. The measure is $\hat{S}(s) = 1$.

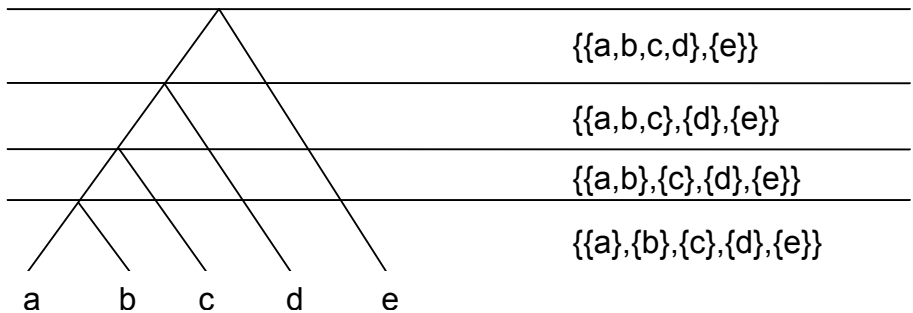

Figure 2.1.2. An ultrametric distance, on the left, and its induced set of partitions, on the right. The vertical distance between the horizontal lines indicates the measure of the corresponding partition.

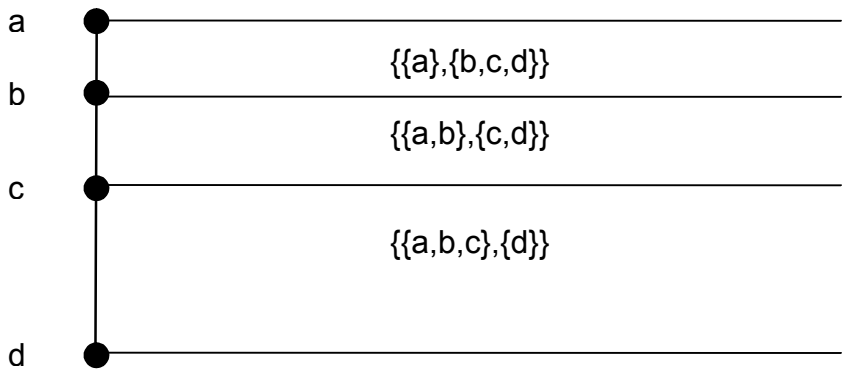

Figure 2.1.3. A set of partitions and a measure on these partitions for states which are points on a real interval (see text for details). The states, {a,b,c,d}, are presented on a vertical interval on the left. On the right appear the sets of partitions induced by the location of states, separated by horizontal lines emerging from the states. The vertical distance between the horizontal lines indicates the measure of the corresponding partition.

---

[8] The ultrametricity of U guarantees that $s_r$ is a partition.
[9] This partition structure is identical to the structure constructed in 1.3.7 after banding.

The reason that partitions are an interesting object of study in IT is the equivalence between partitions and reduced alphabets. Formally,

**Definition 2.1.3.** (reduced alphabets):

Let A and B be two alphabets with probabilities on letters $P_A$ and $P_B$, correspondingly. B is a **reduced alphabet** of A if there is a map $\Psi: A \to B$ of states in A onto states in B so that $\Psi$ preserves the probability, i.e. for any b in B

$$P_B(b) = \sum_{\substack{a \text{ so that } \Psi(a)=b}} P_A(a) = P_A(\Psi^{-1}(b))$$

$\Psi$ is referred to as the **alphabet reducing map**.

**Definition 2.1.4** (partitions as reduced alphabets): Given an alphabet A with probability P on states and s, a partition of A:

1. define for any $a \, \varepsilon \, A$,

    $i_{a,s}$ is the single component of s that contains a.

2. the mapping $i_{a,s}: A \to s$ is an alphabet reducing map when the probability $P_S$ on s is defined as $P_S(j) = P(j)$ for any j in s.

3. This enables us to view the partition s as a **reduced alphabet of A**.

**Statement 2.1.5**: Equivalence of reduced alphabets and partitions: There is a mapping from the class of reduced alphabets of A onto the set of partitions of A.

Thus, it is possible to view partitions and reduced alphabets as identical and interchangeable concepts. Hence, partition structures can be viewed as measures on a set of reduced alphabets.

When studying a set of states A with a partition structure $(S, \hat{S})$ and a probability P on A, another probability (Q) emerges:

**Definition 2.1.6**: Let P be a probability on a <u>finite</u> set of states A, and let $(S, \hat{S})$ be a partition-structure on A. We define

1. $\underline{S} = \{(i,s): i \varepsilon s \varepsilon S\}$.

2. Q is a probability on $\underline{S}$ defined for any $(i,s) \varepsilon \underline{S}$ as: $Q(i,s) = P(i) * \hat{S}(s)$.

    The probability $Q(i,s)$ is the product of the measure $(\hat{S})$ of s and the probability of the letter i in the alphabet s.

**Lemma 2.1.7**: Q is a probability on $\underline{S}$ if and only if $\hat{S}$ is normalized.

Proof: Q is non negative and all we need to check is normalization

$$\sum_{(i,s) \varepsilon \underline{S}} Q(i, s) = \sum_{s \varepsilon S} \sum_{i \varepsilon s} P(i) * \hat{S}(s) = \sum_{s \varepsilon S} \hat{S}(s) \qquad \blacksquare$$

The rest of this section presents some necessary definitions.

**Definition 2.1.8** (Cartesian product of partitions and partition structures): Let $(S_A, \hat{S}_A)$ be a partition-structure on A and let $(S_B, \hat{S}_B)$ be a partition-structure on B. Their Cartesian product is defined so that $(S_A \times S_B, \hat{S}_A \times \hat{S}_B)$ is a partition-structure of AxB:

1. For any $s_A \varepsilon S_A$ and $s_B \varepsilon S_B$ : $s_A \times s_B = \{i \times j : i \varepsilon s_A \text{ and } j \varepsilon s_B\}$ is a partition of AxB.

2. $S_A \times S_B = \{s_A \times s_B : s_A \varepsilon S_A \text{ and } s_B \varepsilon S_B\}$ is a set of partitions of AxB.

3. $\hat{S}_A \times \hat{S}_B (s_A \times s_B) = \hat{S}_A(s_A) * \hat{S}_B(s_B)$, is a measure on the set of partitions $S_A \times S_B$.

**Note 2.1.9**: In 2.1.8.1 $s_A \times s_B$ is a partition of AxB hence it can be interpreted as a reduced alphabet. Read as a reduced alphabet, $s_A \times s_B = \{(i,j) : i \varepsilon s_A \text{ and } j \varepsilon s_B\}$.

**Definition 2.1.10** (probabilities on partitions/reduced-alphabets):

1. Given a probability P on A and a partition s of A (interpreted as a reduced alphabet), the probability $P^S$ on the reduced alphabet s is defined as

$$P^S(i) = \sum_{a \varepsilon i} P(a) = P(i) \qquad \text{for any } i \varepsilon s,$$

When the context permits, we abbreviate $P^S(i)$ to $P(i)$.[10]

2. Assume $P_{AB}$ is a joint probability on $A \times B$. Assume also that $s_A$ and $s_B$ are partitions of A and B, correspondingly. $s_A \times s_B$ is a partition of $A \times B$ which can also be viewed as a reduced alphabet (see note 2.1.9). $P_{AB}^{S_A, S_B}$ is the probability on the reduced alphabet $s_A \times s_B$. It is defined by

$$P_{AB}^{S_A, S_B}(i,j) = P_{AB}(i \times j) = \sum_{a \varepsilon i \, b \varepsilon j} P_{AB}(a, b), \qquad \text{for any } i \varepsilon s_A \text{ and } j \varepsilon s_B.$$

When the context permits, we abbreviate $P_{AB}^{S_A, S_B}(i,j)$ to $P_{AB}(i,j)$ or $P(i,j)$.

3. Assume $P_{AB}$ is a joint probability on $A \times B$ and that $s_A$ and $s_B$ are partitions of A and B, correspondingly. $s_A \times s_B$ is a partition of $A \times B$ which can also be viewed as a reduced alphabet (see note 2.1.9). $P_{AB}^{S_A | S_B}$ will denote the conditional probability derived from $P_{AB}^{S_A, S_B}$.

$$P_{AB}^{S_A | S_B}(i|j) = P_{AB}^{S_A, S_B}(i,j) \, / \sum_{b \varepsilon j} P_{AB}^{S_A, S_B}(i,j) \ , \qquad \text{for any } i \varepsilon s_A \text{ and } j \varepsilon s_B.$$

**Definition 2.1.11** (Combining partition-structures): We say that $\hat{S}_3 = \hat{S}_1 + \hat{S}_2$ if $S_3 = S_1 \cup S_2$ and for any $s \varepsilon S_3$ $\hat{S}_3(s) = \hat{S}_1(s) + \hat{S}_2(s)$ where, on the right hand side, if s is not in the scope then the value should be interpreted as 0. In this definition it is not assumed that the measures are normalized nor that they are separating.

**Definition 2.1.12** (Joining of partitions): If s and t are partitions of A, the partition s∩t is defined by $s \cap t = \{i \cap j : i \varepsilon s, j \varepsilon t \text{ and } i \cap j \neq \phi\}$.

**Definition 2.1.13** (Refinement of partitions): If s and t are partitions of A, the partition s is said to refine t if s∩t=s. In this terminology the partition to singletons refines every partition.

---

[10] One can think of P(i) as the probability of the subset i of A or, equivalently, as the probability of the letter i in the reduced alphabet induced by s.

## Section 2.2 Translation from IT to IT$_S$

To develop a structure sensitive information theory, this section, defines structure sensitive analogs of some notions in IT: entropy, joint entropy, conditional entropy, mutual information and relative entropy. These will be referred to as 'basic notions'. The next section shows that they retain their traditional properties. In following sections other notions will find their structure sensitive counterparts.

First, a definition of structure sensitive entropy (H$_S$) is presented, exactly as stated in 1.3.9. for the ultrametric case. Let P be a probability on A and s a partition of A. When s is read as a reduced alphabet, it has a probability, P$^S$, defined on the letters of the alphabet s (see 2.1.10). P$^S$ has a (traditional) entropy, namely H(P$^S$). H(P$^S$) can be computed for every partition s$\varepsilon$ S and the weighted average of these values can calculated, using the weights of Ŝ(s). This yields (note the similarity to 1.3.9):

$$H_S = \sum_{s\varepsilon S} \hat{S}(s) * H(P^S)$$

This is equivalent to

$$H_S = \sum_{s\varepsilon S} \hat{S}(s) * \sum_{i\varepsilon s} P^S(i) * \log(1/P^S(i)) = \sum_{s\varepsilon S} \sum_{i\varepsilon s} \hat{S}(s) * P^S(i) * \log(1/P^S(i))$$

In anticipation of a formal translation of formulae of the basic notions of IT to their structure-sensitive counterparts, note that such translation must translate H to the H$_S$. Comparing the two expressions

$$H_S = \sum_{s\varepsilon S} \sum_{i\varepsilon s} \hat{S}(s) * P^S(i) * \log(1/P^S(i)) \qquad H = \sum_{a\varepsilon A} P(a) * \log(1/P(a))$$

shows that there are two main differences between H and H$_S$: a) the range of their states is different – for the traditional case the states are A whereas for the structure-sensitive case the states are $\underline{S}$ (see definition 2.1.6). b) in H$_S$ there is weighing by Ŝ which has no counterpart in H.

The translation of the basic notions of IT from their traditional form to their structure-sensitive form will be accomplished via the following translation procedure:

**Definition 2.2.1: The translation procedure for the basic notions.** To translate a basic notion of traditional IT to its structure-sensitive form, proceed as follows:

- Translate any expression of the form P(a) that appears in the scope of a log function to P$^S$(i) and, similarly, any P$_{AB}$(a,b) to P$_{AB}^{S_A,S_B}$(i,j).
- Translate any remaining P(a) (where this expression is not in the scope of a log function) to Ŝ(s) * P$^S$(i). Similarly, P(a,b) needs to be translated to Ŝ(s$_A$,s$_B$) * P$_{AB}^{S_A,S_B}$(i,j).
- Translate $\sum_{a\varepsilon A}$ to $\sum_{s\varepsilon S} \sum_{i\varepsilon s}$ and $\sum_{a\varepsilon A}\sum_{b\varepsilon B}$ to $\sum_{s_A\varepsilon S_A,\, s_B\varepsilon S_B} \sum_{i\varepsilon s_A,\, j\varepsilon s_B}$ .

This translation procedure will now be applied to the basic notions of IT. Each basic notion will have three forms: a. the traditional definition; b. the result of applying the translation procedure (2.2.1); c. another form, in which the structure-sensitive notion is interpreted in terms of reduced alphabets.

Note that all the definitions below will be meaningful even when $\hat{S}$ is not normalized.

**Definition 2.2.2**: **Entropy** is traditionally defined as

2.2.2.a $\qquad H(P) = \sum_{a \varepsilon A} P(a) * \log(1/P(a)).$

The structure sensitive translation is

2.2.2.b $\qquad H_S(A,P,\hat{S}) = \sum_{s \varepsilon S} \hat{S}(s) * \sum_{i \varepsilon S} P^S(i) * \log(1/P^S(i))$

The translation above is equivalent to the following form stated in terms of the entropy of the reduced alphabets

2.2.2.c $\qquad H_S(A,P,\hat{S}) = \sum_{s \varepsilon S} \hat{S}(s) * H(P^S)$

When context permits, $H_S(A,P,\hat{S})$ will be replaced by a shorter notation – $H_S(P,\hat{S})$, $H_S(P)$ or even $H_S$.

**Definition 2.2.3: Entropy of a joint distribution** is traditionally defined as

2.2.3.a $\qquad H(P_{AB}) = \sum_{a \varepsilon A \ b \varepsilon B} P_{AB}(a,b) * \log(1/P_{AB}(a,b)).$

The structure sensitive translation is

2.2.3.b $\qquad H_S(AxB, P_{AB}, \hat{S}_A x \hat{S}_B) = \sum_{s_A \varepsilon S_A \ s_B \varepsilon S_B} \hat{S}(s_A,s_B) * \sum_{i \varepsilon s_A \ j \varepsilon s_B} P_{AB}^{S_A,S_B}(i,j) * \log(1/ P_{AB}^{S_A,S_B}(i,j))$

The translation above is equivalent to the following form stated in terms of the entropy of the joint reduced alphabets

2.2.3.c $\qquad H_S(AxB, P_{AB}, \hat{S}_A x \hat{S}_B) = \sum_{s_A \varepsilon S_A \ s_B \varepsilon S_B} \hat{S}(s_A,s_B) * H(P_{AB}^{S_A,S_B})$

**Definition 2.2.4: Conditional Entropy** is defined when $P_{AB}$ is a joint probability on AxB, $P_B$ is its marginal probability on B and $P_{A|B}$ is the probability on A conditioned on B. Traditionally conditional entropy is defined as

2.2.4.a $\qquad H(P_{A|B}) = \sum_{a \varepsilon A} P_B(b) * \sum_{b \varepsilon B} P_{AB}(a,b)/P_B(b) * \log(P_B(b)/P_{AB}(a,b)) =$

$\qquad\qquad = \sum_{a \varepsilon A \ b \varepsilon B} P_{AB}(a,b) * \log(P_B(b)/P_{AB}(a,b)).$

The structure sensitive translation of the last expression is

2.2.4.b      $H_S(AxB, P_{A|B}, \hat{S}_A x \hat{S}_B) =$

$$= \sum_{s_A \varepsilon S_A \ s_B \varepsilon S_B} \hat{S}(s_A, s_B) * \sum_{i \varepsilon s_A \ j \varepsilon s_B} P_{AB}^{S_A, S_B}(i,j) * \log(P_B^{S_B}(j) / P_{AB}^{S_A, S_B}(i,j))$$

The translation above is equivalent to the following form stated in terms of the conditional entropies of the reduced alphabets. A shorter notation for the left side is adopted and the notation in 2.1.10.3 is used:

2.2.4.c      $H_S(P_{A|B}) = \sum_{s_A \varepsilon S_A \ s_B \varepsilon S_B} \hat{S}(s_A, s_B) * H(P_{AB}^{S_A|S_B})$

**Definition 2.2.5: Mutual Information** is traditionally defined as follows for a joint probability $P_{AB}$ on AxB with marginal probabilities $P_A$ and $P_B$:

2.2.5.a      $I(P_A; P_B) = \sum_{a \varepsilon A \ b \varepsilon B} P(a,b) * \log(P(a,b)/(P(a)*P(b)))$

The structure sensitive translation is (with the same meaning of $P_{AB}$, $P_A$ and $P_B$)

2.2.5.b      $I_S((A, P_A(A), \hat{S}_A) ; (B, P_B(B), \hat{S}_B)) =$

$$= \sum_{s_A \varepsilon S_A \ s_B \varepsilon S_B} \hat{S}_A(s_A) * \hat{S}_B(s_B) * \sum_{i \varepsilon s_A \ j \varepsilon s_B} P_{AB}^{S_A, S_B}(i,j) * \log(P_{AB}^{S_A, S_B}(i,j)/(P_A^{S_A}(i) * P_B^{S_B}(j))$$

The translation above is equivalent to the following form stated in terms of the mutual information of the reduced alphabets. Adopting a shorter notation for the left side leads to:

2.2.5.c      $I_S(P_A; P_B) = \sum_{s_A \varepsilon S_A \ s_B \varepsilon S_B} \hat{S}_A(s_A) * \hat{S}_B(s_B) * I(P_A^{S_A}; P_B^{S_B})$

**Definition 2.2.6: Relative Entropy** is defined for two probabilities on the same set of states. The traditional definition is

2.2.6.a      $D(P_A || P_B) = \sum_{a \varepsilon A} P_A(a) * \log(P_A(a)/P_B(a))$

The structure sensitive translation assumes identical states and an identical structure $\hat{S}$. The translation is

2.2.6.b      $D_S((A, P_A(A), \hat{S}) || (A, P_B(A), \hat{S})) = \sum_{s \varepsilon S} \hat{S}(s) * \sum_{i \varepsilon s} P_A^S(i) * \log(P_A^S(i)/P_B^S(i))$

The translation above is equivalent to the following form stated in terms of the relative information of the reduced alphabets. Adopting a shorter notation for the left side leads to:

2.2.6.c      $D_S(P_A || P_B) = \sum_{s \varepsilon S} \hat{S}(s) * D(P_A^S(A) || P_B^S(A))$

## Section 2.3 Structure sensitive information theory – basic properties

This section begins with showing that some of the well known traditional relations (see (1) chapter 2) are retained in $IT_S$. Then some other properties of basic structure-sensitive notions are presented. The proof of the structure-sensitive analogues of traditional theorems is based on form c. in the definitions 2.2.2-2.2.6 (which is a formulation in terms of reduced alphabets). In this form each notion is presented as a weighted average (weighted according to the measure $\hat{S}$) of the corresponding traditional notion applied to the reduced alphabets induced by the partitions. The fact that the structure sensitive notions are weighted averages of their corresponding traditional notions implies (as shown below) that the basic traditional theorems in IT carry over to $IT_S$ straightforwardly.

**Theorem 2.3.1.   Chain rule for entropy**. In analogy to the traditional chain rule for entropy, $H(P_{AB}) = H(P_A) + H(P_{B|A})$, in $IT_S$ the following holds: If $\hat{S}$ is normalized then

$$H_S(P_{AB}) = H_S(P_A) + H_S(P_{B|A}).$$

Proof: by 2.2.2.c, 2.2.3.c and 2.2.4.c, the claim translates to

$$\sum_{s_A \varepsilon S_A \, s_B \varepsilon S_B} \hat{S}(s_A,s_B) * H(P_{AB}^{S_A,S_B}) = \sum_{s_A \varepsilon S_A} \hat{S}(s_A) * H(P_A^{S_A}) + \sum_{s_A \varepsilon S_A \, s_B \varepsilon S_B} \hat{S}(s_A,s_B) * H(P_{AB}^{S_A|S_B})$$

Since $\hat{S}(s_A,s_B) = \hat{S}(s_A) * \hat{S}(s_B)$ and since $\sum \hat{S}(s_B) = 1$ (by normalization of $\hat{S}$), it suffices to show that for any $s_A, s_B$

$$H(P_{AB}^{S_A,S_B}) = H(P_A^{S_A}) + H(P_{AB}^{S_A|S_B})$$

But this follows immediately from the chain rule for entropy in the traditional IT. ∎

**Theorem 2.3.2. Expressing mutual information by entropy and conditional entropy**.
In traditional IT we have $I(P_A; P_B) = H(P_A) - H(P_{A|B}) = H(P_B) - H(P_{B|A})$. In analogy, in $IT_S$, if $\hat{S}_A$ and $\hat{S}_B$ are normalized then the following holds:

$$I_S(P_A; P_B) = H_S(P_A) - H_S(P_{A|B}) = H_S(P_B) - H_S(P_{B|A}).$$

Proof: To prove       $I_S(P_A; P_B) = H_S(P_A) - H_S(P_{A|B})$,

replace the expressions by their corresponding definitions

$$\sum_{s_A \varepsilon S_A \, s_B \varepsilon S_B} \hat{S}_A(s_A) * \hat{S}_B(s_B) * I(P_A^{S_A}; P_B^{S_B}) = \sum_{s_A \varepsilon S_A} \hat{S}(s_A) * H(P_A^{S_A}) - \sum_{s_A \varepsilon S_A \, s_B \varepsilon S_B} \hat{S}(s_A,s_B) * H(P_{AB}^{S_B|S_A})$$

Since $\hat{S}(s_A,s_B) = \hat{S}(s_A) * \hat{S}(s_B)$ and since $\sum \hat{S}(s_B) = 1$ (by normalization of $\hat{S}$), it suffices to show that for any $s_A, s_B$

$$I(P_A^{S_A}; P_B^{S_B}) = H(P_A^{S_A}) - H(P_{AB}^{S_B|S_A})$$

But this follows immediately from the corresponding property in the traditional IT. ∎

**Theorem 2.3.3. Concavity of entropy**. In analogy to traditional entropy, $H_S$ is a concave function of the probability of each state.

Proof: It suffices to show that the second derivative of $H_S$ is negative.

$$H_S''(P_a) = (\sum_{s \varepsilon S} \hat{S}(s) * H(P^S))'' = (\sum_{s \varepsilon S} \hat{S}(s) * \sum_{i \varepsilon S} P_i * \log(1/P_i))'' =$$

$$= \sum_{s \varepsilon S} \hat{S}(s) * (P_{i_{a,s}} * \log(1/P_{i_{a,s}}))''$$

The result, $H_S''(p_a) \leq 0$, follows from the fact that $\hat{S}(s) \geq 0$, that $P_{i_{a,s}}$ is a linear function of $P_a$ and that $(P * \log(1/P))'' \leq 0$. ∎

**Theorem 2.3.4. Non negativity of entropy, mutual information and relative information**. In analogy to traditional IT, also in $IT_S$: $H_S$, $H_S(A|B)$, $I_S(A;B)$ and $D_S(P_A||P_B)$ are non negative.

Proof: Form c of definitions 2.2.2-2.2.6 presents each structure-sensitive notion as a weighted average (according to $\hat{S}$) of the corresponding traditional notion applied to the reduced alphabets. Since the traditional notions are guaranteed to be non-negative, and since the measure $\hat{S}$ is non-negative, the structure-sensitive counterparts must also be non-negative. ∎

**Theorem 2.3.5**. If $\hat{S}$ is normalized $H_S$ has the following additional formulations, expressed in terms of the probability Q defined in 2.1.6:

1.  $H_S(A, P, \hat{S}) = \sum_{s \varepsilon S} \sum_{i \varepsilon S} Q(i,s) * \log(1/P(i))$
2.  $H_S(A, P, \hat{S}) = H(Q) - H(\hat{S}) = H(Q|\hat{S})$

Proof. 1. is immediate from the definition of $H_S$ (2.2.2) and the definition of Q (2.1.6.). As for 2., the first equation below holds due to traditional grouping and the others follow from the definitions.

$$H(Q) = H(\hat{S}) + \sum_{s \varepsilon S} \hat{S}(s) * H(Q|S=s)$$

$$= H(\hat{S}) + \sum_{s \varepsilon S} \hat{S}(s) * H(P^S) = H(\hat{S}) + H_S$$

Thus $H_S = H(Q) - H(\hat{S}) = H(Q, \hat{S}) - H(\hat{S}) = H(Q|\hat{S})$ ∎

The expression $H_S = H(Q|\hat{S})$ means that $H_S$ is the traditional entropy of Q adjusted for the entropy of the structure $\hat{S}$.

**Observation 2.3.6.** The formulation $H_S = \Sigma \ \hat{S}(s) * H(P^S)$ has the following implications:

1. It is valid even if $\hat{S}$ does not sum to 1. Formally, if $\Sigma \ \hat{S}(s) = C > 0$, one could normalize $\hat{S}$ to $\hat{S}_1 = \hat{S}/C$ and then define $H_S(A,P,\hat{S}) = C * H_S(A,P,\hat{S}_1)$. This is consistent with defining $H_S(A,P,\hat{S}) = \Sigma \ \hat{S}(s) * H(P^S)$. The normalization of $\hat{S}$ is essential for Q to be a probability but otherwise $\hat{S}$ need not be normalized. This was demonstrated in the definitions 2.2.2-2.2.6 where $\hat{S}$ was not required to be normalized and it will be utilized below.

2. Since for any partition s, $H(P^S) \leq H(P)$, and since $H_S = \Sigma \ \hat{S}(s) * H(P^S)$, when $\hat{S}$ is normalized (i.e. $\Sigma \ \hat{S}(s) = 1$), it follows that $H_S \leq H(P)$. Traditional entropy is tantamount to the assumption that S has a single partition to singletons, thus viewing the states merely in their highest level of resolution. Contrary to that, $H_S$ views the states at various levels of resolution, hence $H_S \leq H(P)$.

**Theorem 2.3.7. Additivity in structure.** Entropy, joint entropy, mutual information and relative information are all additive in $\hat{S}$. Thus, for example, if $\hat{S}_3 = \hat{S}_1 + \hat{S}_2$ then $H_{S_3} = H_{S_1} + H_{S_2}$.

Proof: By form c. of definitions 2.2.2-2.2.6 a structure-sensitive notion is equal to a weighted average (according to $\hat{S}$) of the corresponding traditional notion applied to the reduced alphabets. As such, these notions are additive in $\hat{S}$ ∎

So far it has been shown, quite effortlessly, that $IT_S$ satisfies elementary theorems of IT (e.g. the basic theorems in (1) chapter 2). If $IT_S$ is to be taken seriously as an extension of IT, then it needs to have an analogue to the traditional theory beyond that. Hence, the next sections study other aspects of $IT_S$.

**Section 2.4 Typical sequences in $IT_S$**

In this section $H_S$ is interpreted in terms of the number and probability of typical sequences. In doing so several kinds of sequences will be discussed, using the following terminology:

**Definition 2.4.1.** When referring to sequences of length N drawn IID from a known probability the following terminology will be used:
  1. $\underline{S}$-sequences: sequences over the alphabet $\underline{S}$ with probability Q (see 2.1.6)
  2. S-sequences: sequences over the alphabet S with probability $\hat{S}$ (for normalized $\hat{S}$)
  3. A-sequences: sequences over the alphabet A with probability P
  4. s-sequences: sequences on the alphabet of s (where s is a partition of A) with probability $P^S$ (see definition 2.1.10)

The sequence length will be added as a prefix, if needed. Thus a m-A-sequence is a sequence of length m over the alphabet A with probability P.

A $\underline{S}$-sequence can be uniquely projected on a S-sequence. For a big enough N, typical N-$\underline{S}$-sequences have a projection which is a typical S-sequence. The fact that $H_s = H(Q) - H(\hat{S})$ implies that the $\sim 2^{NH(Q)}$ typical N-$\underline{S}$-sequences can be partitioned to $\sim 2^{NH(\hat{S})}$ equivalence classes where in each class all projections on S are the same. Every equivalence class has the same cardinality: $\sim 2^{N*H_S}$. This is one manner in which $N*H_S$ appears as the cardinality of typical sequences.

The statement in the previous paragraph is the structure-sensitive analogue of the traditional theorem stating that the number of typical sequences is $\sim 2^{N*H}$. To fill in the details of this analogy, note that in the traditional case S has a single partition to singletons, hence the projection on S yields a constant result and $H(\hat{S})=0$. Thus, in the traditional case, since all $\underline{S}$-typical sequences are in the same equivalence class and since $\underline{S}$-sequences are equivalent to A-sequences – the number of typical sequences is $2^{N*H}= 2^{N*H_S} = 2^{N*H(Q)}$.

In traditional IT there is another interpretation of entropy in terms of typical sequences: the probability of a typical sequence is $\sim 2^{-N*H}$. The analogue of this statement for $IT_S$ will now be presented. For a traditional typical sequence $(x_1,\ldots,x_N)$,

$$p(x_1,\ldots,x_N) = \prod_{j=1,\ldots,N} p(x_j) = \sim \prod_{i=1,\ldots n} (p_i)^{n*p_i} = 2^{-N*H}.$$

Consider a typical N-$\underline{S}$-sequence $(x_1,\ldots,x_N)$ and interpret its symbols as subsets of A. Each such subset has a probability according to P, to which we shall refer as $P(x_i)$. It is easy to show that $\prod P(x_i) = \sim 2^{-N*H_S}$. Thus, $\sim 2^{-N*H_S}$ is the P-probability of a typical N-$\underline{S}$-sequence when interpreting its symbols as subsets of A. This yields another interpretation of $H_S$ in terms of probability of typical sequences. Again, when applied to the traditional case this reduces to the well known traditional statement that the probability of a typical sequence is $\sim 2^{-N*H}$.

Up till now the typical sequences that were discussed were on the alphabet $\underline{S}$ and not on the alphabet A. In the next sections an interpretation of $N*H_S$ in terms of typical N-A-sequences will be sought. This will be meaningful in discussing coding.

## Section 2.5  The grouping property in $IT_S$ and the concordance distance

The subject of this section is the form that the grouping property has in $IT_S$.

The traditional grouping property is a most meaningful axiom in the definition of entropy in traditional IT. It states that for any partition s

$$H(P) = H(P^s) + \sum_{i \in s} P(i)*H(P|i) = H(P^s) + H(P|s).$$

The traditional grouping property cannot hold for $H_S$ since this would imply that $H_S=H$. We do know that the chain rule for entropy does hold for $H_S$ (2.3.1), so where does the grouping property fail for $H_S$? Before proceeding there is the need for some notation.

**Definition 2.5.1** (notation on restricting partitions to a subset): Given $\hat{S}$, a measure on a set S of partitions of A and any B⊆A

1. For s∈S, the restriction of s to B is s|B = {i∩B : i ε s}.
2. Define the restriction of S to B as S|B = {s|B : s∈S and s|B is a partition of B}.
3. For any u∈S, if u|B is a partition of B, let $S_u$={s∈S: so that s|B=u|B}
4. $\hat{S}$|B is the measure on S|B defined as

$$(\hat{S}|B)(s|B) = \sum_{t \text{ in } S_s} \hat{S}(t)$$

**Note 2.5.2**: Given S, a set of partitions of A, a subset B⊆A and a partition s∈S.

1. The cardinality of s|B might be smaller than the cardinality of s since a component of s might become empty when restricted to B. Hence s|B might not be a partition (e.g. when s={B,A−B}). s|B is a partition of B iff |s|B| > 1 (its cardinality exceeds 1).
2. The set S|B is a set of partitions of B: S|B={s|B:s∈S and |s|B|>1}. The cardinality of S|B might be smaller than the cardinality of S since two different partitions of A might become identical when restricted to B and since some partitions of A might stop being a partition when restricted to B.
3. Starting with a normalized $\hat{S}$, the measure $\hat{S}$|B might not be normalized.

To examine the grouping property in $IT_s$ let's start with a simple case: let A be an alphabet with probability P, let $\hat{S}$ be a normalized partition structure composed of a single partition s (i.e. $\hat{S}$=(s,1)) and let t be a binary partition of A. The analogy to the traditional grouping is

$$H_s(A, P, (s,1)) = H_s(t, P^t, (t,1)) + \sum_{i \text{ in } t} P(i) * H_s(i, P|i, (s|i,1))$$

Since $H_s(A,P,(s,1))$=$H(P^s)$ and $H_s(t,P^t,(t,1))$= $H(P^t)$, the equality above is equivalent to

$$H(P^s) = H(P^t) + \sum_{i \text{ in } t} P(i) * H_s(i, P|i, (s|i), 1).$$

When does this hold? Due to the structure-sensitive chain rule (2.3.1), this holds only when s refines t (i.e. s = s ∩ t see 2.1.13).[11] But what if t does not refine s?

It can be shown that if s and t are independent (i.e. for any i∈s, j∈t P(i|j)=P(i)∗P(j)) then

$$H(P^s) = \sum_{i \text{ in } t} P(i) * H_s(i, P|i, (s|i), 1).$$

These two examples indicate that the grouping depends on the relation between s and t. The following lemma describes some relations between $H(P^s)$, $H(P^t)$ and $H(P^{s \cap t})$.

**Lemma 2.5.3.** It is easy to see that for any partition s:

1. $H(P^s) \le H(P^{s \cap t}) \le H(P^s) + H(P^t)$,
2. $0 \le [H(P^s) - H(P^{s \cap t}) + H(P^t)]$,
3. $0 \le [H(P^s) - H(P^{s \cap t}) + H(P^t)]/H(P^t) \le 1$.

Proof: immediate. ∎

---

[11] This is similar to (1.3.2) in $IT_u$ where the grouping-like property holds only for the natural partition.

Based on 2.5.3., the following theorem establishes the grouping for the case $\hat{S}=(s,1)$:

**Theorem 2.5.4.** (grouping in a simple case): Assume P is a probability on A and s is a partition of A. Let $t=\{A_1,A_2\}$ be a binary partition of A and let the concordance of t and s be

$$C(t,s) = 1/H(P^t) * [H(P^s) - H(P^{s\cap t}) + H(P^t)]$$

then   $H(P^s) = C(t,s) * H(P^t) + \sum_{i \text{ in } t} P(i) * H_s(i,P|i,(s|i,1))$

Proof: The proof appears in the Appendix.

Additivity in structure (see 2.3.7) suggests how to move from $\hat{S}=(s,1)$ to any structure $\hat{S}$:

**Definition 2.5.5.** (concordance and distance between complementary subsets): Assume P is a probability on A and $\hat{S}$ is a measure on partitions of A. Let $t=\{A_1,A_2\}$ be a binary partition of A . We define

1.  the **concordance** between t and $\hat{S}$ is

$$C(t, \hat{S}) = 1/H(P^t) * \sum_{s \text{ in } S} \hat{S}(s) * [H(P^s) - H(P^{s\cap t}) + H(P^t)].$$

2.  the **concordance distance** between the complementary sets $A_1$ and $A_2$ is[12]

$$d_{\hat{S}}(A_1,A_2) = C_{\hat{S}}(t, \hat{S})$$

The concordance distance is additive in the structure $\hat{S}$:

**Theorem 2.5.6.** $d_{\hat{S}}(i_1,i_2)$ is additive in $\hat{S}$:  if $\hat{S}_3 = \hat{S}_1 + \hat{S}_2$ then $d_{\hat{S}_3} = d_{\hat{S}_1} + d_{\hat{S}_2}$.

Proof: Immediate from definition 2.5.5.2∎

Using the concordance distance (2.5.5) and its additivity (2.5.6) it is possible to formulate and prove structure-sensitive grouping:

**Theorem 2.5.7.** (grouping): Assume P is a probability on A and $\hat{S}$ is a measure on partitions of A. Let $t=\{A_1,A_2\}$ be a binary partition of A then

$$H_S(A,P,\hat{S}) = d_{\hat{S}}(A_1,A_2) * H(P^t) + \sum_{i \text{ in } t} P(i) * H_S(i,P|i,\hat{S}|i).$$

Proof: See the Appendix.

In traditional IT, $d_{\hat{S}} \equiv 1$. This can be seen from 2.5.5.: the traditional partition structure (see 2.1.2.3) has a single partition, s, composed of singletons, so for any t, $s=s\cap t$, so $H(P^s)=H(P^{s\cap t})$ and $d_{\hat{S}}(A_1,A_2)=1/H(P^t) * \hat{S}(s) * H(P^t)=1$.

In the next two sections we learn more about $d_{\hat{S}}$ and its role in coding.

---

[12]   The way things were presented, $d_s$ is defined only on complementary sets of states. However this is not really a restriction and one can require merely that they are disjoint (since one can condition the probability and the structure on $A_1 \cup A_2$ and within this, $A_1$ and $A_2$ are complementary sets).

**Section 2.6  The concordance distance between sets of letters: $d_{\hat{S}}$**

In this section we will study $d_{\hat{S}}$. The next theorem presents another formulation of $d_{\hat{S}}$ which will be useful when studying coding.

**Theorem 2.6.1.** (another formulation of $d_{\hat{S}}$): Assume P is a probability on A and $\hat{S}$ is a measure on partitions of A. Let t={$A_1,A_2$} be a binary partition of A. then

$$d_{\hat{S}}(A_1,A_2) = [H_{\hat{S}}(A, P, \hat{S}) - \sum_{j \in t} P(j) * H_{\hat{S}}(j,P|j,\hat{S}|j)] / H(P^t)$$

$$= [H_{\hat{S}}(A, P, \hat{S}) - \sum_{j=1,2} P(A_j) * H_{\hat{S}}(A_j,P|A_j,\hat{S}|A_j)] / H(P^t) \ .$$

Proof: By the additivity of $d_{\hat{S}}$ in $\hat{S}$ (2.5.6), it suffices to prove the theorem for the case of a single partition. The proof is in the Appendix.

**Theorem 2.6.2.**
1. if $\hat{S}$ includes a single partition s
   a. $0 \le d_{\hat{S}}(A_1,A_2) \le \hat{S}(s)$.
   b. if $H(P^s)=H(P^{s \cap t})$ then $d_{\hat{S}}(A_1,A_2)= \hat{S}(s)$.
2. $0 \le d_{\hat{S}}(A_1,A_2) \le \sum_{s \text{ in } S} \hat{S}(s)$.

Proof: immediate from 2.6.1 and 2.5.3. ∎

**Conclusion 2.6.3.** (non negativity of $d_{\hat{S}}$): Let P be a probability on A, $\hat{S}$ be a measure on partitions of A and let $i_1,i_2$ be two complementary non-null subsets of A. Then $d_{\hat{S}}(i_1,i_2)$ is non negative.

Proof: that is 2.6.2.2. ∎

$d_{\hat{S}}(i_1,i_2)$ is defined for any two complementary subsets, $i_1,i_2$. However, if the partition t={$i_1,i_2$} is itself in S, a stronger result than 2.6.3 can be stated for the value of $d_{\hat{S}}(i_1,i_2)$:

**Conclusion 2.6.4.** Let P be a probability on A, $\hat{S}$ be a measure on partitions of A and let t={$i_1,i_2$} be a partition in S. Then $d_{\hat{S}}(i_1,i_2) \ge \hat{S}$ (t).

Proof: $d_{\hat{S}}(A_1,A_2) = 1/H(P^t) * \sum_{s \text{ in } S} \hat{S}(s) * [H(P^s)-H(P^{s \cap t})+ H(P^t)] \ge$

$$1/H(P^t) * \hat{S}(t) * [H(P^t)-H(P^{t \cap t})+ H(P^t)] = \hat{S}(t). \ ∎$$

**Definition 2.6.5**: (distance between states):

$$D_{\hat{S}}(a,b) = \sum_{s \text{ separates } a \text{ from } b} \hat{S}(s)$$

will be referred to as the **distance between states induced by $\hat{S}$**.

**Theorem 2.6.6**. Let P be a probability on A and Ŝ be a measure on partitions of A.

1. $D_{\hat{S}}$, defined for singletons, is a distance.
2. $D_{\hat{S}}$ is identical to $d_{\hat{S}}$ when the latter is restricted to singletons.

Proof. The only non-obvious property required to prove 1. is that $D_{\hat{S}}$ satisfies the triangle inequality. This results from the fact that

$$\{s : i_{a,s} \neq i_{c,s}\} \subseteq \{s : i_{a,s} \neq i_{b,s}\} \cup \{s : i_{b,s} \neq i_{c,s}\}$$

To prove 2., apply $d_{\hat{S}}$ to two singletons and let t={{a},{b}}

$$d_{\hat{S}}(\{a\},\{b\}) = 1/H(P^t) * \sum_{s \text{ in } S} \hat{S}(s) * [H(P^s) - H(P^{s \cap t}) + H(P^t)] =$$

$$d_{\hat{S}}(\{a\},\{b\}) = 1/H(P^t) * \sum_{s \text{ separates a from b}} \hat{S}(s) * [H(P^s) - H(P^{s \cap t}) + H(P^t)] =$$

$$d_{\hat{S}}(\{a\},\{b\}) = \sum_{s \text{ separates a from b}} \hat{S}(s) \quad . \qquad \blacksquare$$

The next section discusses coding and compression and shows that $d_{\hat{S}}$ is the key to interpreting $H_S$ as the lower bound of compression.

## Section 2.7  Coding in IT$_S$

Given a code tree, what is the effect of the structure Ŝ on coding? The answer ties code length with $d_{\hat{S}}$, in line with section 1.4, as follows:

**Definition 2.7.1**. Let C be a binary code tree with leaves A, P a probability on A and a structure Ŝ.

1. For each node i in the code tree let $A_i$ = {a: a is a leaf under i}

2. For each node i in the code tree let {$A_{i,L}$, $A_{i,R}$} be the partition of $A_i$ to two sets of leaves according to the branching of the code tree in the node i.

3. For each node i in the code tree let $d^i$ be $d_{\hat{S}}(A_{i,L}, A_{i,R})$.

4. For any leaf a, define its code length

$$CL(a) = \sum_{i \text{ is a node on the path from the root to a, } i \neq a} d^i \quad .$$

5. The expected structure-sensitive code length (ESSCL) is the expected value of code lengths, i.e.

$$ESSCL(C,A,P,\hat{S}) = \sum_{a \text{ is a leaf}} P(a) * CL(a) = \sum_{i \text{ is a non leaf node}} P(A_i) * d^i \quad .$$

Definition 2.7.1 formalizes the effect of the structure Ŝ on coding. Each node in the code tree is assigned a merit according to the distance between the two sets of letters that the node separates (2.7.1.3). In traditional IT the code length of a letter is the number of nodes leading from the root of the code tree to the letter (leaf). The structure-sensitive code length of a letter is calculated in a similar manner but each node in the path does not increase the length by 1 but rather by its merit (2.7.1.4). The average code length is computed from the code lengths of the letters in the same way as in traditional IT (2.7.1.5)

In analogy to traditional IT:

**Theorem 2.7.2**. Given an alphabet A with probability P and a normalized structure Ŝ, $H_S \leq ESSCL(C,A,P,\hat{S})$.

Proof: Take a binary code tree C and use its structure to define a hierarchy of partitions. The grouping rule applies to each of the partitions.

Using the notation of 2.7.1, continued application of the grouping rule according to the code tree leads to:

$$H_S = \sum P(A_i) * d^i * H(P(A_{i,L}), P(A_{i,R}))$$
$$\text{i is a non-leaf node in the code tree}$$

By definition 2.7.1.5:

$$ESSCL(C,A,P,\hat{S}) = \sum P(A_i) * d^i * 1$$
$$\text{i is a non-leaf node in the code tree}$$

Since $H(P(A_{i,L}), P(A_{i,R}))$ is the entropy of two states it cannot exceed 1 hence $H_S$ cannot exceed the expected code-length. ∎

In traditional IT, when taking long enough A-sequences generated IID via P, it is possible to devise a code on the m-A-sequences with expected code length as near to $m*H$ as wanted. The structure sensitive analogue to this theorem involves a code tree where:

1. The alphabet is $A^m$, i.e. the leaves of the tree are of m-A-sequences.

2. The probability of a leaf, a sequence $a_1, …, a_m$, is $P^m(a_1, …, a_m) = \prod P(a_i)$,

3. The partition-structure of $A^m$ is $\hat{S}^m$ (i.e. $\hat{S} X \hat{S} X … X \hat{S}$ m times as defined in 2.1.8, see also 2.1.9). The structure $\hat{S}^m$ on $A^m$ is composed of partitions $s_1 x … x s_m$ where $s_i \varepsilon S$. The partition $s_1 x … x s_m$ is defined (see 2.1.8.1) so that $a_1, …, a_m$ and $b_1, …, b_m$ are in the same component iff for every $1 \leq i \leq m$, $a_i$ and $b_i$ are in the same component of $s_i$.

**Example 2.7.3**: Here are examples of the structure $\hat{S}^m$ on $A^m$.

1. Assume Ŝ is the traditional structure composed of a single partition to singletons that has a measure of 1. $\hat{S}^m$ has a single partition of $A^m$ to singletons (i.e. m-A-sequences), again with measure 1.

2. Let A = {a,b,c,d}, S = { {{a},{b},{c},{d}}, {{a,b},{c,d}} } and the measure Ŝ of the partition is .6 & .4 correspondingly. $\hat{S}^2$ partitions $A^2$ as follows (the measures appear on the left):

.36   { {(a,a)}, {(a,b)}, {(a,c)}, {(a,d)}, {(b,a)}, {(b,b)}, {(b,c)}, {(b,d)},
      {(c,a)}, {(c,b)}, {(c,c)}, {(c,d)}, {(d,a)}, {(d,b)}, {(d,c)}, {(d,d)} }

.24   { {(a,a),(a,b)}, {(a,c), (a,d)}, {(b,a), (b,b)}, {(b,c), (b,d)},
      {(c,a), (c,b)}, {(c,c), (c,d)}, {(d,a), (d,b)}, {(d,c), (d,d)} }

.24   { {(a,a), (b,a)}, {(a,b), (b,b)}, {(a,c), (b,c)}, {(a,d), (b,d)},
      {(c,a), (d,a)}, {(c,b), (d,b)}, {(c,c), (d,c)}, {(c,d), (d,d)} }

.16   { {(a,a), (a,b), (b,a), (b,b)}, {(a,c),(a,d),(b,c),(b,d)},
      {(c,a), (c,b), (d,a), (d,b)}, {(c,c), (c,d), (d,c), (d,d)} }

In analogy to traditional IT:

**Theorem 2.7.4**. Given an alphabet A with probability P and a normalized structure Ŝ and $\varepsilon > 0$, it is possible to find a big enough N so that for any $m > N$ m-A-sequences can be encoded in a code tree C with

$$\text{ESSCL}(C, A^m, P^m, \hat{S}^m) \le m * (H_S(A, P, \hat{S}) + \varepsilon).$$

Proof: The theorem can be proved using essentially the same code tree used in traditional IT. See Appendix.

## Section 2.8 – Extending IT to a structure sensitive IT$_S$

IT applies to any alphabet and any probability on the letters of the alphabet. It tells us how to handle combinations of alphabets and how to handle reduced alphabets. Therefore, it has a way of addressing a combination of reduced alphabets, which is the structure Ŝ introduced in section 2.1. In this sense, IT$_S$ is an application within traditional IT.

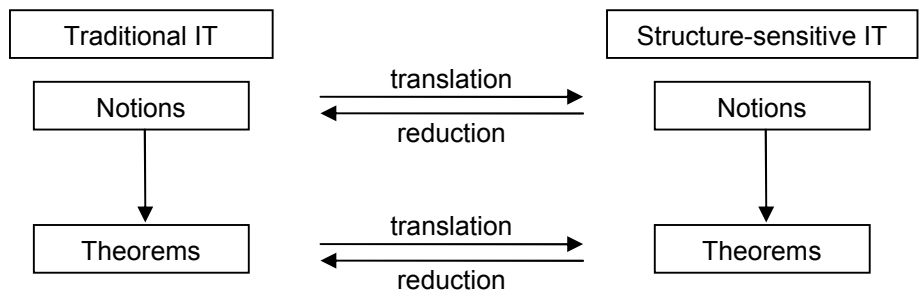

Figure 2.8.1. IT$_S$ generalizes IT: an analogy between structure-sensitive IT and traditional IT. There is a translation process from traditional notions to their structure-sensitive analogs. This enables translating a traditional theorem to its structure-sensitive counterpart. The structure-sensitive theorems are true. The structure-sensitive notions and theorems reduce back to their traditional counterparts when interpreted in the case a traditional structure.

However things can be viewed differently: although IT$_S$ is within IT it generalizes IT. The generalization has a formal meaning presented in Figure 2.8.1. It states that it is possible to translate notions of traditional IT to their structure-sensitive counterparts so that the translation of theorems yields valid theorems in IT$_S$. This was proven for the basic notions and their elementary properties (sections 2.2 and 2.3), using an explicit translation procedure. Section 2.4 on typical sequences, section 2.5 on grouping and section 2.7 on coding and compression – all have a similar nature: they all show, that notions are translatable so that validity of theorems is retained. Contrary to sections 2.2 and 2.3, in sections 2.4, 2.5 and 2.7 no formal translation procedure was presented since there are only a few formal results [13]. Further study is required to check how the extendibility portrayed in Figure 2.8.1 applies to other areas of IT not treated in this manuscript.

Figure 2.8.1 also shows backward reduction, in the following sense. The traditional case is tantamount to a structure with a single partition (to singletons), see 2.1.2.3. We will refer to that as the traditional structure. When a notion or a theorem of $IT_S$ is restricted to the traditional structure, it reduces back the traditional notion or theorem. Thus, $IT_S$ with the traditional structure is the traditional IT. Hence $IT_S$ can be viewed not merely as an extension of traditional IT but also as a generalization of it (or IT).

---

[13]  Without an explicit translation procedure, the translation is error prone since the way things are translated is essential to retaining truth of theorems. For example, the first compression theorem fails in IT$_S$ if the traditional notion of code length is retained – only if the two notions of entropy and of code length are translated, then the compression theorem holds in IT$_S$.

The claim that $IT_S$ is a generalization of IT has two implications on the foundations of information theory:

a. It demonstrates that the assumption of IT that there is no structure on the underlying alphabet is not necessary: IT can cope with a structure that is expressed in appropriate terms.

b. It suggests a conceptual framework for expressing structure of the underlying alphabet. The intuitive concept of structure can be expressed formally in numerous ways. The fact that $IT_S$ is a full fledged information theory suggests that a measure on a set of partitions is an appropriate framework to express structure for IT, since it retains the fundamentals of IT.

$IT_S$ has practical implications: it extends the scope of cases to which IT can be rigorously applied. To give a few examples:

1. We saw (section 1.1 and application 1.5.1) that to measure positional protein conservation within the framework of IT, a reduced alphabet has been suggested so as to accommodate the implications of an underlying structure. $IT_S$ offers a richer repertoire of alternatives. There is still the option to continue using a single reduced alphabet, but other options emerge: adopting a hierarchical structure, adopting a linear structure (see Part 3), or even a Cartesian product of any of those.

2. In 1.5.2 we saw that the spin glass model requires computation of entropy when the states are ultrametric. The formulation used there was $H_U$ for a two-layer hierarchical structure. Applying an unrestricted formulation of $H_U$ might provide a more accurate formulation as well as a formulation that is more continuous in phase transition.

3. In applications of IT to analyze real-valued random variables, there are numerous situations where the data is binned and the varying levels of similarity between bins is disregarded. That can be handled more conveniently and accurately within $IT_S$, as will be presented in Part 3.

$IT_S$ does not exclude the possibility that there are other approaches of incorporating structure in IT. Moreover, there is no claim that given a relation on states, formulating it as a partition structure captures the essence of the relation. Neither is there a claim that the choice of partitions to convey a given relation is necessarily unique. Considerations external to $IT_S$ are required to assess whether a certain partition structure properly conveys an underlying relation. External considerations are also required to choose the most appropriate partition structure in cases where more than one such structure seems suited. Examples of such external considerations are:

1. In Part 1 and Part 3 there is a natural distance between states and the partition structure is derived from that distance (see 2.1.2.1 and 2.1.2.2). Interestingly, the converse is also true – the partition structure retains sufficient information to reconstruct the natural distance: the distance defined in 2.6.5 is the natural distance from which the structure was derived. In this sense the partition structure captures all the information that there is in the distance and could be viewed as conveying the natural distance sufficiently.

2. Part 1 and Part 3 present intuitive desiderata regarding the function expressing entropy. Desiderata of a similar nature might be possible in other contexts and not merely regarding the notion of entropy.

Since $IT_S$ is all about a weighted combination of reduced alphabets, and since the usage of reduced alphabets is completely within IT, $IT_S$ is completely within traditional IT. The reason for emphasizing the unavoidably controversial view of $IT_S$ being a generalization of IT is that its implications on the foundations of IT and on its range of applications is clearer.

$IT_S$ applies to any case where structure is expressed as a measure on partitions. If the nature of the partitions is restricted in one way or another, some unique features might emerge. Consider the restriction to hierarchical sets of partitions (i.e. where for any two partitions one is the refinement of the other), which leads to $IT_U$ (see Part 1). Since the traditional partition is hierarchical the following relations hold (where < indicates reduction of scope):

$$IT < IT_U < IT_S$$

Part 3 restricts $IT_S$ to another set of partition-structures, very different from hierarchical partitions.

# Part 3 – extending IT to accommodate distances in R (IT$_R$)

## Section 3.1 IT$_R$ as a special case of IT$_S$

Part 2 showed that a structure-sensitive extension of IT can be constructed given a measure on partitions of A. This will be applied to the case where the states are real numbers. Such states appear in numerous applications of IT where there is an underlying (real valued) random variable, e.g. levels of grey of a pixel, biochemical properties of amino acids and value of a stock, just to mention a few.

**Definition 3.1.1.** Assume A is a finite set of real numbers: $A=\{a_i\}_{i=1,\ldots,n}$, where $a_i < a_j$, for $i<j$. Assume that on A there is a probability P, denoted as $\{P(a_i)\}_{i=1,\ldots,n}$ or $\{P_i\}_{i=1,\ldots,n}$. (see Figure 3.1.1.)

1. A is normalized if $a_n - a_1 = 1$. Unless explicitly stated normalization will not be assumed.

2. For any $i<n$ let $s_i^-$ be $\{a_j: j \leq i\}$ and let $s_i^+$ be $\{a_j: j>i\}$.

3. For any $i<n$ let $s_i$ be the partition of A defined as $s_i = \{s_i^-, s_i^+\}$.

4. Define the following partition structure: $S = \{s_i : i<n\}$ and $\hat{S}(s_i) = a_{i+1} - a_i$.

5. The structure sensitive IT resulting from restricting structures to the pairs $(S,\hat{S})$ above will be referred to as IT$_R$.

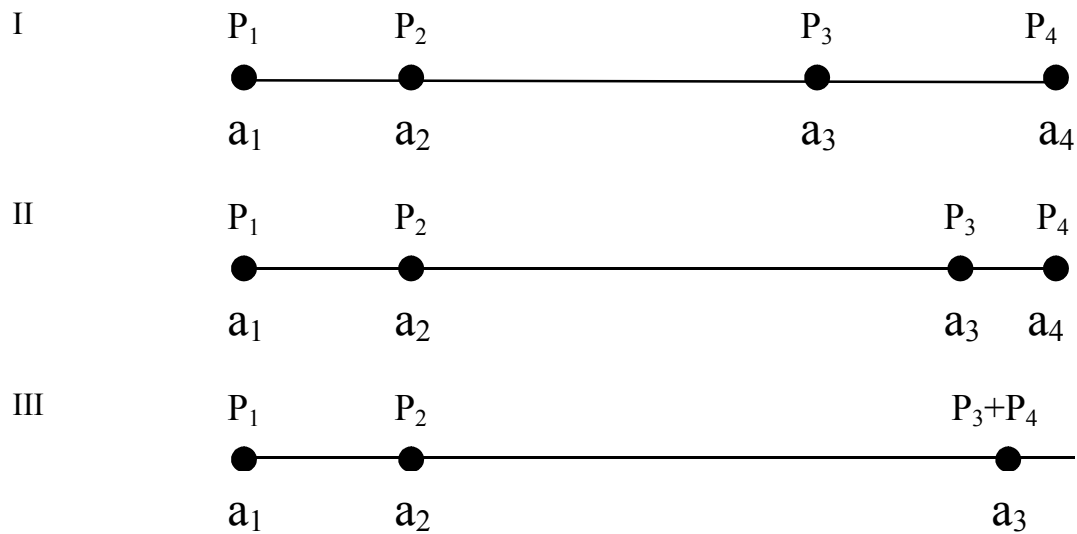

Figure 3.1.1. I. The objects of our interest are points on a real line on which a probability is defined. II. The states $a_3$ & $a_4$ are very near and the entropy of the system should be similar to that of III.

Definition 3.1.1 reflects the intention is to develop an IT for probabilities on real numbers by using the results of Part 2 thus establishing IT$_R$ as a special case of IT$_S$. To accomplish that – a set of partitions, S, with a measure $\hat{S}$ were defined. However, given a finite set of real numbers, there is no unique manner of defining $(S,\hat{S})$ and there is no a-priori guarantee that the partition structure defined in 3.1.1 captures the relation between states properly (see the discussion in 2.8). There is no guarantee of properly capturing the distance even though the distance between states can be reconstructed from Ŝ. Hence, in order to be able to assess whether the to-be-developed-$IT_R$ captures the essence of the structure, we start with setting some expectations stated in terms of the, to be constructed, structure-sensitive entropy $H_R$.

**Expectations 3.1.2**. The to-be-defined $H_R$ is a measure of 'randomness' resembling variance. It is expected to satisfy the following inequalities (just as an example)[14]:

1. Consider the example presented in Figure 3.1.II. Due to the vicinity of $a_3$ & $a_4$ $H_R$ in Figure 3.1.II should be nearer to that of 3.1.III than to that of 3.1.I, as long as $a_3$ is sufficiently near to $a_4$ in 3.1.II .

2. for any $\varepsilon \leq \frac{1}{3}$: $H_R (\{0, \frac{1}{2}, 1\}, \{\frac{1}{3}, \frac{1}{3} - \varepsilon, \frac{1}{3} + \varepsilon\}) > H_R (\{0, \frac{1}{2}, 1\}, \{\frac{1}{3}, \frac{1}{3} + \varepsilon, \frac{1}{3} - \varepsilon\})$

3. for any $\varepsilon \leq \frac{1}{2}$: $H_R (\{0, \frac{1}{2} - \varepsilon, 1\}, \{\frac{1}{4}, \frac{1}{4}, \frac{1}{2}\}) > H_R (\{0, \frac{1}{2} + \varepsilon, 1\}, \{\frac{1}{4}, \frac{1}{4}, \frac{1}{2}\})$

**Note 3.1.3**. Notes on definition 3.1.1.

1. S does not depend in any way on the probability P. It depends only on the set of real numbers A.

2. S is in 1-1 correspondence with the set of intervals $\{(a_i, a_{i+1}): i<n\}$ and the measure of the partition corresponding to $(a_i, a_{i+1})$ is the length of the interval.

## Section 3.2 The formulation of basic notions for $IT_R$

The implications of $IT_R$ being a structure sensitive IT are that the findings presented in Part 2 apply. Since the structure is more specific than that presented in Part 2, the translation of the basic notions can be stated more specifically. In any case, the basic theorems hold, see 2.3.1-2.3.4, and the results in $IT_S$ regarding typical sequences, coding and compression apply to $IT_R$ and $d_Ŝ$ all hold.

The basic notions in $IT_R$ are:

**3.2.1 Entropy**: Let $A=\{a_i\}_{i=1,...,n}$ be an ordered sets of real numbers and let P be a probability on A . The structure sensitive entropy is defined as

$$H_R (A,P,D) = \sum_{i < n} (a_{i+1} - a_i) * H(P^{\{s_i^-, s_i^+\}}) = \sum_{i < n} (a_{i+1} - a_i) * H(P(s_i^-), P(s_i^+)) =$$

$$= \sum_{i < n} (a_{i+1} - a_i) * [P(s_i^-) * \log(1/ P(s_i^-)) + P(s_i^+) * \log(1/ P(s_i^+))]$$

**3.2.2 Cartesian product**: Let $A=\{a_i\}_{i=1,...,n}$ and $B=\{b_i\}_{i=1,...,m}$ be ordered sets of real numbers and let P be a joint probability on $A \times B$.[15] Let $\{s_{A,i}:i<n\}$ and $\{s_{B,j}:j<m\}$ denote the binary partitions of A and B correspondingly. $H_R (A \times B, P)$ is defined as

$$H_R(A \times B, P) = \sum_{i < n} (a_{i+1} - a_i) * \sum_{j < m} (b_{k+1} - b_k) * H(\{s_{A,i}^{-}, s_{A,i}^{+}\} \times \{s_{B,j}^{-}, s_{B,j}^{+}\}, P)$$

---

[14] It is easy to check that these expectations are met.
[15] Note that the Cartesian product A X B has a distance defined on each component but not on the product set.

For a normalized set A $0 \leq H_R(A, P) \leq 1$. However for a Cartesian product of two normalized sets A and B, $0 \leq H_R(A \times B, P_{AB}) \leq 2$. This is in line with the traditional inequality $H(A \times B, P_{AB}) \leq H(A, P_A) + H(B, P_B)$.

**3.2.3 Conditional probability**: Let A=$\{a_i\}_{i=1,\ldots,n}$ and B=$\{b_i\}_{i=1,\ldots,m}$ be ordered sets of real numbers and let P be a joint probability on $A \times B$. The conditional entropy is defined as:

$$H_R(B \mid A) = \sum_{i < n, j = +-1} Q(i, j) * H_R(B \mid A^*(i, j))$$

**3.2.4 Mutual Information**. Let A=$\{a_i\}_{i=1,\ldots,n}$ and B=$\{b_i\}_{i=1,\ldots,m}$ be ordered sets of real numbers and let P be a joint probability on $A \times B$. Let $\{s_{A,i}:i<n\}$ and $\{s_{B,j}:j<m\}$ denote the binary partitions of A and B correspondingly. Then

$$I_R(A ; B) = \sum_{i < n} \sum_{j < m} (a_{i+1} - a_i) * (b_{j+1} - b_j) * I((P(s_{A,i}^{-}), P(s_{A,i}^{+})); (P(s_{B,j}^{-}), P(s_{B,j}^{+})))$$

**3.2.5 Relative Entropy**. Let A=$\{a_i\}_{i=1,\ldots,n}$ be an ordered sets of real numbers and let $P^1, P^2$ be probabilities on A. Then

$$DKL_R(P^1 \| P^2) = \sum_{i < n} (a_{i+1} - a_i) * DKL((P^1(s_i^{-}), P^1(s_i^{+})) \| (P^2(s_i^{-}), P^2(s_i^{+})))$$

**3.2.6 Entropy of a sample**. In some applications of $IT_R$ there is a (real valued) random variable from which a finite sample is drawn. In such case every sampled value is unique and we have a set A=$\{a_i\}_{i=1,\ldots,n}$, where $a_i < a_j$, for $i<j$. For such case, the entropy is

$$H_R(A, P, D) = \sum_{0 < i < n} (a_{i+1} - a_i) * [i/n * \log(n/i) + (1 - i/n) * \log(n/(n-i))].$$

Similar formulations exist for all other basic notions.

## Section 3.3 Discussing $IT_R$

$H_R$ is relevant in contexts where quantization is performed. Given a (real) random variable with cumulative distribution $F_x$, quantization breaks the distribution into a finite set of bins on which $H_R$ can be calculated. Like the value of H, the value of $H_R$ is clearly dependent on the choice of bins. However, the dependence of $H_R$ on the choice of bins is not so strong since there is a limit when bins become infinitesimal: $\int H(F_x, 1-F_x) dx$, where H(p,1-p) is the standard entropy on two states. $H_R$ is different from the differential entropy: differential entropy can be negative and this limit (like $H_R$) cannot.

Interestingly, simulations show a high level of resemblance between $H_R$ and standard deviation: drawing random samples of points from a uniform distribution and calculating their $H_R$ and their standard deviation yielded a Pearson correlation >0.95 between the two measures of dispersion. A similar correlation resulted when drawing points from a normal distribution. Thus $H_R$ can be viewed as the information-theoretic equivalent of standard-deviation.

The basic machinery of $IT_R$ can be applied not only to distances in R but also to Cartesian products of such sets. However, this does not imply that $IT_R$ is applicable to $R^n$ in the sense that it captures properly a distance in $R^n$. Particularly, the extension does not apply to the Euclidean distance as $IT_R$ is sensitive to the coordinate system. For example, points on a diagonal straight line in $R^2$ will not have the same $H_R$ as in the topologically-equivalent case that they are in R.

# Appendix

**Theorem 1.3.4**: Given an ultrametric distance D on a finite set A and a probability P on A, let T be the tree induced by D. Then

[∗]     $H_U(P, D) = \sum\limits_{i \text{ is a non-leaf node in } T} P_i * \text{height}(i) * H(P^{Y_i})$

**Proof**: The proof is by recursion on the ultrametric tree.

If the tree has only one non-leaf node, which by definition must be the root, then part b of the recursive definition applies and $H_U(P, D) = d * H(P)$. This is exactly what [∗] states since $P(A_{root}) = 1$, $\text{height}(root) = d$ and $P^{Y_{root}} = 1$.

Otherwise, let $Y = \{A_i : i \text{ is a direct descendant of the root}\}$ be the natural-partition of A. From part a of the definition and using the recursive assumption that [∗] holds for all sub-trees of the ultrametric tree, we have that for any node $t$, which is a direct descendant of the root,

$H_U(P|A_i, D|A_i) = \sum\limits_{j \text{ is a non-leaf descendant of } i} (P|A_i)_j * \text{height}(j) * H(P^{Y_j})$

Since $P(A_i) * (P|A_i)_j = P(A_j)$, we have

$P(A_i) * H_U(P|A_i, D|A_i) = \sum\limits_{j \text{ is a non-leaf descendant of } i} P(A_j) * \text{height}(j) * H(P^{Y_j})$

therefore, using Lemma 1.3.3 and the recursive definition of $H_U$

$H_U(P, D) = H(P^Y) * \text{height}(root) + \sum\limits_{i \text{ is a non-leaf node that is not the root}} P(A_i) * \text{height}(i) * H(P^{Y_i})$

$= \sum\limits_{i \text{ is a non-leaf node}} P(A_i) * \text{height}(i) * H(P^{Y_i})$

which proves [∗]. ∎

**Theorem 1.3.5**: Given an ultrametric distance D on a finite set A and a probability P on A, let T be the tree induced by D.

Then         $H_U(P,D) = - \sum\limits_{i \text{ is a non-root node in } T} L_i * P_i * \log(P_i)$

**Proof**: To prove the theorem, it suffices to show that

$- \sum\limits_{k \text{ is a non-root node}} L_k * P_k * \log(P_k) = \sum\limits_{i \text{ is a non-leaf node}} P_i * \text{height}(i) * H(P^{Y_i})$

For every non-leaf node **i**,

$$P_i * \text{height}(i) * H(P^{Y_i}) = -P_i * \text{height}(i) * (\sum_{j \text{ is a direct descendant of } i} (P_j / P_i) * \log(P_j / P_i))$$

$$= -\text{height}(i) * (\sum_{j \text{ is a direct descendant of } i} P_j * \log(P_j / P_i))$$

$$= -\text{height}(i) * (\sum_{j \text{ is a direct descendant of } t} P_j * \log(P_j) - P_i * \log(P_i)) =$$

$$= \text{height}(i) * P_i * \log(P_i) - \sum_{j \text{ is a direct descendant of } i} \text{height}(i) * P_j * \log(P_j).$$

Summing across all non-leaf nodes **i** we get

$$\sum_{i \text{ is a non-leaf node}} P_i * \text{height}(i) * H(P^{Y_i}) =$$

$$\sum_{i \text{ is a non-leaf node}} \text{height}(i) * P_i * \log(P_i) - \sum_{j \text{ is a non-root node}} \text{height}(\text{Parent}(j)) * P_j * \log(P_j) =$$

$$= -\sum_{j \text{ is a non-root node}} (\text{height}(\text{Parent}(j)) - \text{height}(j)) * P_j * \log(P_j) = -\sum_{j \text{ is a non-root node}} L_j * P_j * \log(P_j)$$

∎

**Theorem 1.3.9.**:

$$H_U = \sum_{t \text{ is realizable}} \hat{S}(t) * H(s_t)$$

**Proof**: By lemma 1.3.8 it is possible to assume that T is banded. By 1.3.5

$$H_U(P,D) = -\sum_{i \text{ is a non-root node in T}} L_i * P_i * \log(P_i)$$

$$= -\sum_{t \text{ is realizable}} \sum_{i \text{ is in } r_t} L_i * P_i * \log(P_i)$$

$$= -\sum_{t \text{ is realizable}} \hat{S}(t) \sum_{i \text{ is in } r_t} P_i * \log(P_i)$$

$$= -\sum_{t \text{ is realizable}} \hat{S}(t) * H(s_t) \qquad ∎$$

The following theorem is needed later when discussing compression.

**Theorem 1.3.10** [16] (Minimality of $H_u$ versus arbitrary binary partitions): Let $Y=\{A_1,A_2\}$ be any binary partition of states (not necessarily the natural partition), then

**[1]**     $H_u(P, D) \leq H_u(P^Y, D^Y) + \sum_{j=1,2} P(A_j) * H_u(P|A_j, D|A_j)$

Comments:

- In this rather lengthy proof we refer to several contexts in which $H_u$ is used. In some cases, in order to stress the underlying alphabet, we will use the form $H_u(A, P, D)$ instead of $H_u(P, D)$. For example, note that the alphabet underlying the computation of $H_u(P^Y, D^Y)$ is Y so we will sometimes refer to $H_u(P^Y, D^Y)$ as $H_u(Y, P^Y, D^Y)$.
- Since $H_u$ was defined only for ultrametric distances, for the statement **[1]** to be meaningful it must be confirmed that $H_u(Y, P^Y, D^Y)$ is well defined. Since Y has a two-letter alphabet - any metric on it is ultrametric. [17]
- Note that

    $H_u(P^Y, D^Y) = D(A_1,A_2) * h(P(A_1)) =$

    $D(A_1,A_2) * [P(A_1)*\log(P(A_1)) + P(A_2)*\log(P(A_2))]$

**Proof**: The proof is by induction on the cardinality of the alphabet.

We use the following notation:

   $Z=\{A^1,\ldots, A^n\}$ is the natural partition of the alphabet (according to D)

   $A_j{}^i = A_j \cap A^i$ for $j=1,2$ and $i=1,\ldots,n$

   $Y^i = \{A_1{}^i, A_2{}^i\}$ is a partition of $A^i$ for $i=1,\ldots,n$. Note that some $A_j{}^i$ might be void.

   $Z_j = \{A_j{}^1, \ldots, A_j{}^n\}$ is a partition of $A_j$ for $j=1,2$. Note that some $A_j{}^i$ might be void.

If we denote the maximal distance of D by d then for any $1 \leq \kappa, L \leq n$, $D(A^K, A^L) = d$ and for any $1 \leq j \leq 2$ if $A_j{}^K$ and $A_j{}^L$ are not null then $D(A_j{}^K, A_j{}^L) = d$.

| | | $A^1$ | $A^2$ | ...... | $A^n$ | Row partition Z - natural |
|---|---|---|---|---|---|---|
| | $A_1$ | $A_1{}^1$ | $A_1{}^2$ | | $A_1{}^n$ | $Z_1$ |
| | $A_2$ | $A_2{}^1$ | $A_2{}^2$ | | $A_2{}^n$ | $Z_2$ |
| Column partition | Y | $Y^1$ | $Y^2$ | | $Y^n$ | |

Table 1: auxiliary table for Theorem A.

---

[16] This theorem can be proved by using theorem 2.5.7. However, a direct proof is provided here.
[17] This is the reason the claim of miniality is made solely for binary partitions: if the partition is to more than two, then $H_u(P_Y, D_Y)$ might not be defined.

Since

$$H_u(A, P, D) = H_u(Z, P^Z, D^Z) + \sum_{i=1,n} P(A^i) * H_u(A^i, P|A^i, D|A^i) \quad \text{and}$$

$$H_u(A_j, P|A_j, D|A_j) = H_u(Z_j, P^{z_j}, D^{z_j}) + \sum_{i=1,n} P(A_j^i|A_j) * H_u(A_j^i, P|A_j^i, D|A_j^i)$$

proving **[1]** is equivalent to proving **[2]**

**[2]** $\quad H_u(Z, P^Z, D^Z) + \sum_{i=1,n} P(A^i) * H_u(A^i, P|A^i, D|A^i) \quad \leq$

$$H_u(P^Y, D^Y) + \sum_{j=1,2} P(A_j) * \left[ H_u(Z_j, P^{z_j}, D^{z_j}) + \sum_{i=1,n} P(A_j^i|A_j) * H_u(A_j^i, P|A_j^i, D|A_j^i) \right] =$$

$$H_u(P^Y, D^Y) + \sum_{j=1,2} P(A_j) * H_u(Z_j, P^{z_j}, D^{z_j}) + \sum_{i=1,n} \sum_{j=1,2} P(A_j^i) * H_u(A_j^i, P|A_j^i, D|A_j^i)$$

by the induction hypothesis, applied to $A^i$, we know that

**[3]** $\quad H_u(A^i, P|A^i, D|A^i) \leq H_u(Y^i, P^{y_i}, D^{y_i}) + \sum_{j=1,2} P(A_j^i|A^i) * H_u(A_j^i, P|A_j^i, D|A_j^i) =$

$$= H_u(Y^i, P^{y_i}, D^{y_i}) + \sum_{j=1,2} P(A_j^i) / P(A^i) * H_u(A_j^i, P|A_j^i, D|A_j^i)$$

from which we derive

**[4]** $\quad \left[ H_u(A^i, P|A^i, D|A^i) - H_u(Y^i, P^{y_i}, D^{y_i}) \right] * P(A^i) \leq \sum_{j=1,2} P(A_j^i) * H_u(A_j^i, P|A_j^i, D|A_j^i)$

therefore, to prove **[2]** it suffices to prove

**[5]** $\quad H_u(Z, P^Z, D^Z) + \sum_{i=1,n} P(A^i) * H_u(A^i, P|A^i, D|A^i) \leq$

$$H_u(P^Y, D^Y) + \sum_{j=1,2} P(A_j) * H_u(Z_j, P^{z_j}, D^{z_j}) +$$

$$\sum_{i=1,n} \left[ H_u(A^i, P|A^i, D|A^i) - H_u(Y^i, P^{y_i}, D^{y_i}) \right] * P(A^i)$$

Which, by elimination of $\sum_{i=1,n} P(A^i) * H_u(A^i, P|A^i, D|A^i)$ from both sides, is equivalent to

**[6]** $\quad H_u(Z, P^Z, D^Z) \leq$

$$H_u(P^Y, D^Y) + \sum_{j=1,2} P(A_j) * H_u(Z_j, P^{z_j}, D^{z_j}) - \sum_{i=1,n} H_u(Y^i, P^{y_i}, D^{y_i}) * P(A^i)$$

Which, is equivalent to

[7]    $H_U(Z, P^Z, D^Z) + \sum_{i=1,n} H_U(Y^i, P^{Yi}, D^{Yi})*P(A^i) \leq H_U(P^Y, D^Y) + \sum_{j=1,2} P(A_j)* H_U(Z_j, P^{zj}, D^{zj})$

Note that in all appearances of $H_U$ in the previous expression, $H_U$ can be replaced by $\sum p*\log(p)*D$. After removing the minus sign and changing the direction of the inequality, we need to prove:

[8]    $\sum_{i=1,n} P(A^i)*\log(P(A^i))*d + \sum_{i=1,n} P(A^i) * \sum_{j=1,2} P(A_j^i| A^i)*\log(P(A_j^i| A^i))*D(A_1^{\ i}, A_2^{\ i}) \geq$

$\sum_{j=1,2} P(A_j)*\log(P(A_j))*D(A_1,A_2) + \sum_{j=1,2} P(A_j)* \sum_{i=1,n} P(A_j^{\ i}|A_j)*\log(P(A_j^{\ i}|A_j))*d$

after some manipulation we get the following equivalent inequality:

[9]    $\sum_{i=1,n} P(A^i)*\log(P(A^i))*d + \sum_{i=1,n} \sum_{j=1,2} P(A_j^{\ i})*\log(P(A_j^{\ i}))* D(A_1^{\ i}, A_2^{\ i})$

$- \sum_{i=1,n} P(A^i)*\log(P(A^i))* D(A_1^{\ i}, A_2^{\ i}) \geq$

$\sum_{j=1,2} P(A_j)*\log(P(A_j))*D(A_1,A_2) + \sum_{j=1,2} \sum_{i=1,n} P(A_j^{\ i})*\log(P(A_j^{\ i}))*d$

$- \sum_{j=1,2} P(A_j)*\log(P(A_j))*d$

or equivalently:

[10]    $\sum_{i=1,n} P(A^i)*\log(P(A^i))*[d - D(A_1^{\ i}, A_2^{\ i})]$

$+ \sum_{i=1,n} \sum_{j=1,2} P(A_j^{\ i})*\log(P(A_j^{\ i}))* [D(A_1^{\ i}, A_2^{\ i}) - d]$

$\geq \sum_{j=1,2} P(A_j)*\log(P(A_j))*[D(A_1,A_2) - d]$

or equivalently:

[11]    $\sum_{i=1,n} [d - D(A_1^{\ i}, A_2^{\ i})] * [P(A^i)*\log(P(A^i)) - \sum_{j=1,2} P(A_j^{\ i})*\log(P(A_j^{\ i}))]$

$\geq [D(A_1,A_2) - d] * \sum_{j=1,2} P(A_j)*\log(P(A_j))$

or equivalently (after some manipulation of the left side of the inequality):

[12]   $\sum_{i=1,n} \left[ D(A_1{}^i, A_2{}^i) - d \right] * \sum_{j=1,2} P(A_j{}^i) * \log(P(A_j{}^i | A^i))$

   $\geq \left[ D(A_1, A_2) - d \right] * \sum_{j=1,2} P(A_j) * \log(P(A_j))$

now we will express $D(A_1, A_2)$ as a function of $D(A_1{}^i, A_2{}^i)$, and use it to replace $D(A_1, A_2)$ in [12]. By definition of distances as the weighted averages of the distances between underlying leaves, we have

[13]   $D(A_1, A_2) = \left[ \sum_{i=1,n} P(A_1{}^i) * ( P(A_2{}^i) * D(A_1{}^i, A_2{}^i) + (P(A_2) - P(A_2{}^i)) * d) \right] / \left[ P(A_1) * P(A_2) \right]$

   $= \left[ \sum_{i=1,n} P(A_1{}^i) * P(A_2{}^i) * D(A_1{}^i, A_2{}^i) \right] / \left[ P(A_1) * P(A_2) \right]$

   $+ \left[ \sum_{i=1,n} P(A_1{}^i) * P(A_2) * d - \sum_{i=1,n} P(A_1{}^i) * P(A_2{}^i) * d \right] / \left[ P(A_1) * P(A_2) \right] =$

   $\left[ \sum_{i=1,n} P(A_1{}^i) * P(A_2{}^i) * (D(A_1{}^i, A_2{}^i) - d) \right] / \left[ P(A_1) * P(A_2) \right] + d$

and therefore

[14]   $D(A_1, A_2) - d = \left[ \sum_{i=1,n} P(A_1{}^i) * P(A_2{}^i) * (D(A_1{}^i, A_2{}^i) - d) \right] / \left[ P(A_1) * P(A_2) \right]$

replacing [14] in [12] we get the following equivalent inequality requiring a proof:

[15]   $\sum_{i=1,n} \left[ D(A_1{}^i, A_2{}^i) - d \right] * \sum_{j=1,2} P(A_j{}^i) * \log(P(A_j{}^i | A^i))$

   $\geq \sum_{i=1,n} \left[ D(A_1{}^i, A_2{}^i) - d \right] P(A_1{}^i) * P(A_2{}^i) / (P(A_1) * P(A_2)) * \sum_{j=1,2} P(A_j) * \log(P(A_j))$

Note that the expressions on both sides of the inequality above have the form:

   $\sum_{i=1,n} \left[ D(A_1{}^i, A_2{}^i) - d \right] . * X_i .$

We will show that the inequality holds for each summand. Since $D(A_1{}^i, A_2{}^i) - d \leq 0$, we need to change the direction of the inequality and show that for any **i**:

[16]   $\sum_{j=1,2} P(A_j{}^i) * \log(P(A_j{}^i | A^i))$

$$\leq P(A_1{}^i) * P(A_2{}^i) / (P(A_1) * P(A_2)) * \sum_{j=1,2} P(A_j) * \log(P(A_j))$$

the above is equivalent to:

[17] $\sum_{j=1,2} P(A_j{}^i) * \left[\log(P(A_j{}^i) - \log(P(A^i))\right] =$

$\left[P(A_1{}^i) * \log(P(A_1{}^i)) + P(A_2{}^i) * \log(P(A_2{}^i))\right] - \left[P(A_1{}^i) + P(A_2{}^i)\right] * \log(P(A_1{}^i) + P(A_2{}^i))$

$\leq P(A_1{}^i) * P(A_2{}^i) / (P(A_1) * P(A_2)) * \left[ P(A_1) * \log(P(A_1)) + P(A_2) * \log(P(A_2))\right]$

or equivalent to:

[18] $\left[\log(P(A_1{}^i)) / P(A_2{}^i) + \log(P(A_2{}^i)) / P(A_1{}^i)\right]$

$- \left[P(A_1{}^i) + P(A_2{}^i)\right] / \left[ P(A_1{}^i) * P(A_2{}^i)\right] * \log(P(A_1{}^i) + P(A_2{}^i))$

$\leq \log(P(A_1)) / P(A_2) + \log(P(A_2)) / P(A_1) - 0 =$

$\log(P(A_1)) / P(A_2) + \log(P(A_2)) / P(A_1) -$

$\left[P(A_1) + P(A_2)\right] / \left[P(A_1) * P(A_2)\right] * \log (P(A_1) + P(A_2))$

Both sides of the inequality have the form

$$F(x,y) = \log(x) / y + \log(y) / x - (x + y) / (x * y) * \log(x + y) ,$$

i.e. the inequality is equivalent to :

[19] $F(P(A_1{}^i), P(A_2{}^i)) \leq F(P(A_1), P(A_2))$

The proof will be completed by showing that for $0 \leq x, y$ , F is monotone increasing in both x and y.

$F'_x = 1 / (x * y) - \log(y) / x^2 - [x * y - y / (x + y)] / (x * y)^2 * \log(x+y) - (x + y) / [x * y * (x + y)]$

$= [x * y - y^2 * \log(y) + y^2 * \log (x + y) - x * y] / [x^2 * y^2]$

$= [\log (x + y) - \log (y) ] / x^2 \geq 0$

Since F is symmetric in x and y, the same applies to $F'_y$, thus showing that F is monotone increasing in both x and y.

We have thus completed the proof, showing that the natural-partition derived from the ultrametric distance D yields the **smallest** possible value for $H_u$, compared to any **binary** partition. ∎

**Theorem 1.4.3.**: For any ultrametric D and binary code tree C on an alphabet A:
$$H_U \leq \mu^c{}_U.$$

**Proof**. The proof is broken down to a series of lemmas.

**Lemma 1.4.3.1**: $\mu^c{}_U$ and $\lambda^c{}_U$ have the following non-recursive formulations. For any node $\mathfrak{c}$ in the binary code tree C, let $A^{\mathfrak{c}}$ be the letters (leaves) falling under $\mathfrak{c}$. $A^{\mathfrak{c}}$ can be partitioned $A^{\mathfrak{c}_0}$ and $A^{\mathfrak{c}_1}$ according to whether the codeword $\mathfrak{c}$ is extended with 0 or with 1. Then

$$\mu^c{}_U(A, P, D) = \sum_{\mathfrak{c} \text{ is a non-leaf node in the binary code tree}} P\,(A^{\mathfrak{c}}) * D\,(A^{\mathfrak{c}_0}, A^{\mathfrak{c}_1})$$

$$\lambda^c{}_U(A, P, D) = \sum_{\mathfrak{c} \text{ is a non-leaf node in the binary code tree}} P\,(A^{\mathfrak{c}}) * D\,(A^{\mathfrak{c}_0}, A^{\mathfrak{c}_1}) * h\,(A^{\mathfrak{c}_0}, A^{\mathfrak{c}_1})$$

**Proof**: Immediate by recursion on C. ∎

**Conclusion 1.4.3.2**: For any distance D and **binary** code tree C
$$\lambda^c{}_U(A, P, D) \leq \mu^c{}_U(A, P, D)$$
**Proof**: This results from lemma 1.4.3.1 and the fact that $h(A_0, A_1) \leq 1$. ∎

**Conclusion 1.4.3.3**: In case D is Hamming, $\mu^c{}_U = \mu^c$.

**Proof**: From 1.4.3.1, $\mu^c{}_U(A, P, D) = \sum_{\mathfrak{c} \text{ is a non-leaf node C}} P\,(A^{\mathfrak{c}}) * D\,(A^{\mathfrak{c}_0}, A^{\mathfrak{c}_1}) = \sum_{\mathfrak{c}} P\,(A^{\mathfrak{c}}) = \mu^c$. ∎

**Lemma 1.4.3.4.**: For any uniform (ultrametric) $D \equiv d$, probability P and binary code tree C on an alphabet A: $\lambda^c{}_U(A, P, D) = d * H$.

**Proof**: From 1.4.3.1, $\lambda^c{}_U(A, P, D) = \sum_{\mathfrak{c} \text{ is a non-leaf node in the binary code tree}} P\,(A^{\mathfrak{c}}) * D\,(A^{\mathfrak{c}_0}, A^{\mathfrak{c}_1}) * h\,(A^{\mathfrak{c}_0}, A^{\mathfrak{c}_1}) =$

$$= d * \sum_{\mathfrak{c} \text{ is a non-leaf node in the binary code tree}} P\,(A^{\mathfrak{c}}) * h\,(A^{\mathfrak{c}_0}, A^{\mathfrak{c}_1}) =$$

$$= d * H$$

The last equality results from a repeated application of the chain rule for entropy, recursively on the code tree. ∎

**Lemma 1.4.3.5.**: For any ultrametric D, probability P and binary code tree C on an alphabet A: $H_U(P, D) \leq \lambda^c{}_U$.

**Proof**: Recursively on the code tree.

If the code tree has two letters then the topologies of the code tree and the ultrametric tree are identical and we have $\lambda^c{}_U = d * H = H_U$, as shown previously.

Now we turn to the recursive step. The binary code tree C has two sub-trees stemming from the root - $C_0$, $C_1$, with underlying letters $A_0$, $A_1$ which define a partition $Y=\{A_0, A_1\}$ of A.

$\lambda^c{}_U = D (A_0,A_1) * h(A_0,A_1) + \sum_{i=0,1} P(A_i) * \lambda^{ci}{}_U (A_i, P|A_i)$  by definition

$\geq D (A_0,A_1) * h(A_0,A_1) + \sum_{i=0,1} P(A_i) * H_U(P|A_i, D|A_i)$  by the recursion assumption

$= H_U(P^Y, D^Y) + P(A_i) * H_U(P|A_i, D|A_i)$  since Y has two letters

$\geq H_U P, D)$  by theorem 1.3.10

∎

**Proof of Theorem 1.4.3**: from lemma 1.4.3.5 and conclusion 1.4.3.2 we have

$H_U \leq \lambda^c{}_U \leq \mu^c{}_U.$   ∎

**Algorithm 1.4.4 Compressed coding algorithm for $H_U$**. There is an algorithm that given an ultrametric distance with leaves A and a probability P on A, produces a binary code tree with $\mu^c{}_U \leq H_U +1$.

Here we describe an algorithm that yields a code of length $\leq H_U +1$

The algorithm Optimize (code tree, distance tree) is called with two arguments – a code tree and an ultrametric distance tree. It returns a revised (optimized) code tree. In recursive calls to this algorithm the distance tree does not change except, possibly, for being restricted to a subtree. Initially the algorithm is called with a code tree that is identical to the ultrametric tree.

Step 1:  Optimization. Optimize each of the two code trees stemming from the root of the input code tree. The result of optimizing the left branch of the input code tree is called the left-code and the result of optimizing the right branch is the right-code.

Step 2:  Unification. A unified code tree needs to be constructed from the (optimized) left-code and right-code. Each of these has two main branches yielding altogether four binary sub-codes. The algorithm seeks all possible ways of cross combining these four sub-codes and chooses the one with the lowest average code length. If the simple combination of the left-code and right-code is optimal the algorithm terminates and returns. Otherwise (i.e. mixing of the left-code with the right-code is optimal) the whole process of optimization is started over again with the mixed code tree as input.

This algorithm converges in spite of its loopy nature in step 2. The reason is that every iteration decreases the average code length. Since there are only a finite number of code trees, the algorithm must converge.

The algorithm was run on a million random combinations of probability and normalized binary ultrametric distances, where the number of states was between 3 and 50. It faultlessly yielded a code tree C so that $\mu^c_u \leq H_u + 1$. We, have no formal analytic proof of that.

**Theorem 2.5.4.** (grouping in a simple case): Assume P is a probability on A and s is a partition of A. Let t={$A_1$,$A_2$} be a binary partition of A and let

$$C(t,s) = 1/H(P^t) * [H(P^s) - H(P^{s \cap t}) + H(P^t)]$$

then

$$H(P^s) = C(t,s) * H(P^t) + \sum_{i \text{ in } t} P(i) * H_s(i,P|i,(s|i,1))$$

Proof: We need to show that

$$H(P^s) = 1/H(P^t) * [H(P^s) - H(P^{s \cap t}) + H(P^t)] * H(P^t) + \sum_{i \text{ in } t} P(i) * H_s(i,P|i,(s|i,1)),$$

which is equivalent to

$$H(P^s) = H(P^s) - H(P^{s \cap t}) + H(P^t) + \sum_{i \text{ in } t} P(i) * H_s(i,P|i,(s|i,1))$$

and that is equivalent to

$$H(P^{s \cap t}) = H(P^t) + \sum_{i \text{ in } t} P(i) * H((P|i)^{s|i})$$

which follows from traditional grouping. ∎

**Theorem 2.5.7.** (grouping): Assume P is a probability on A and Ŝ is a measure on partitions of A. Let t={$A_1$,$A_2$} be a binary partition of A then

$$H_s(A,P,\hat{S}) = d_{\hat{S}}(A_1,A_2) * H(P^t) + \sum_{i \text{ in } t} P(i) * H_s(i,P|i,\hat{S}|i)$$

Proof: Due to the additivity in structure of $H_s$ (see 2.3.7), and due to the aditivty in structure of $d_{\hat{S}}$ (see 2.5.6) it suffices to prove the theorem when the partition structure has a single partition with measure 1, i.e. Ŝ=(s,1). So it is necessary to prove that

$$H_s(A,P,(s,1)) = C_{\hat{S}}(t,(s,1)) * H(P^t) + \sum_{i \text{ in } t} P(i) * H_s(i,P|i,\hat{S}|i).$$

This is equivalent to

$$H_s(A,P,(s,1)) = C(t,s) * H(P^t) + \sum_{i \text{ in } t} P(i) * H_s(i,P|i,\hat{S}|i),$$

and that is exactly 2.5.4. ∎

**Theorem 2.6.1.** Assume P is a probability on A and Ŝ is a measure on partitions of A. Let t={$t_1$,$t_2$} be a binary partition of A. Then

$$d_{\hat{S}}(t_1,t_2) = [H_s(A, P, \hat{S}) - \sum_{j=1,2} P(A_i) * H_s(t_j,P|t_j,\hat{S}|t_j)] / H(P^t)$$

Proof: By the additivity of $d_{\hat{S}}$ in Ŝ (2.5.6), it suffices to prove for the case of a single partition, i.e. S has a single partition s.

By definition 2.5.5.2,

$$d_{\hat{S}}(t_1, t_2) = \hat{S}(s)/H(P^t)*[H(P^s) - H(P^{s \cap t}) + H(P^t)]$$

$$= \hat{S}(s)/H(P^t)*[\sum_{i \text{ in } s}P(A_i)*\log(1/P(A_i)) - \sum_{j=1,2}\sum_{i \text{ in } s}P(A_i \cap t_j)*\log(1/P(A_i \cap t_j)) + \sum_{j=1,2}P(A_j)*\log(1/P(A_j))]$$

$$= [\hat{S}(s)*\sum_{i \text{ in } s}P(A_i)*\log(1/P(A_i)) - \hat{S}(s)*\sum_{j=1,2}P(t_j)*\sum_{i \text{ in } s}P(A_i \cap t_j)/P(t_j)*\log(P(t_j)/P(A_i \cap t_j))] / H(P^t) =$$

$$[H_S(A, P, \hat{S}) - \sum_{j=1,2}P(t_j)*H_S(t_j, P|t_j, \hat{S}|t_j)] / H(P^t) \blacksquare$$

**Theorem 2.7.4**. Given an alphabet A with probability P and a normalized structure $\hat{S}$ and $\varepsilon > 0$, it is possible to find a big enough N so that for m>N m-A-sequences can be encoded with

$$ESSCL \le m*(H_S(A,P,\hat{S}) + \varepsilon).$$

Proof:

Denote by $P^N$ the probability on N-A-sequences resulting from an IID assumption. We know (2.3.1) that $H_S(A^N, P^N, \hat{S}^N) = N*H_S(A, P, \hat{S})$. Moreover, if $\underline{A^N}$ denotes the typical N-A-sequences, then $H_S(A^N, P^N, \hat{S}^N) = H_S(\underline{A^N}, P^N, \hat{S}^N)$. So we need to construct a code tree C on $\underline{A^N}$ so that $ESSCL(C) = H_S(\underline{A^N}, P^N, \hat{S}^N)$. Take C to be any binary balanced code tree for the $2^{N*H(P)}$ typical N-A-sequences. We claim that at any node i in this tree, if the partition establishes by the node is denoted by $t = \{A_{i,L}, A_{i,R}\}$ then

$$H_S(A_i, P|A_i, \hat{S}|A_i) = d_{\hat{S}}(A_{i,L}, A_{i,R}) + \sum_{j \text{ in } t}(P|A_i)(j)*H_S(j, P|j, \hat{S}|j)$$

This follows immediately from 2.5.7:

$$H_S(A_i, P|A_i, \hat{S}|A_i) = d_{\hat{S}}(A_{i,L}, A_{i,R})*H(P^t|A_i) + \sum_{j \text{ in } t}(P|A_i)(j)*H_S(j, P|j, \hat{S}|j)$$

and from the fact that the leaves are equiprobable and the tree is balanced, therefore

$$H(P^t|A_i) = 1.$$

Returning to the notation $d_{\hat{S}}(A_{i,L}, A_{i,R}) = d^i$, the recursion in the formula above can be removed and

$$H_S(A, P, \hat{S}) = \sum_{i \text{ is a non leaf node}}P(A_i)*d^i$$

But this is exactly ESSCL, see 2.7.1.5. $\blacksquare$

## Acknowledgements

We are grateful to the following for fruitful discussions and valuable comments: Hanah Margalit, Benjamin Weiss, Nati Linial, Naftali Tishby, Nir Friedman, Yossi Rinot, Guy Sella and Or Sattath. We are also grateful to Ido Kanter, Eytan Domany and Ofer Biham for clarifying the calculation of entropy in the ultrametric spin glass model.